%% file: paper_arxiv.tex
\documentclass[a4paper,reqno]{amsart}
\usepackage{enumerate,natbib,textcomp,graphicx,amsmath,amssymb,multirow,listings}
\usepackage[ruled]{algorithm2e}
\DeclareMathOperator{\argmin}{argmin}

\DeclareMathOperator{\rank}{rank}

\begin{document}
\title[Exploring Covid-19 Spatiotemporal Dynamics]{\bf Exploring Covid-19 Spatiotemporal Dynamics: \\ Non-Euclidean Spatially Aware Functional Registration}
\author[L. A. Barratt \& J. A. D. Aston]{Luke A. Barratt (\texttt{lab85@cam.ac.uk})\hspace{.2cm} and John A. D. Aston \\
    Statistical Laboratory, DPMMS, University of Cambridge, UK}
\date{}
\maketitle

\bigskip

\begin{abstract}
When it came to Covid-19, timing was everything. This paper considers the spatiotemporal dynamics of the Covid-19 pandemic via a developed methodology of non-Euclidean spatially aware functional registration. In particular, the daily SARS-CoV-2 incidence in each of 380 local authorities in the UK from March to June 2020 is analysed to understand the phase variation of the waves when considered as curves. This is achieved by adapting a traditional registration method (that of local variation analysis) to account for the clear spatial dependencies in the data. This adapted methodology is shown via simulation studies to perform substantially better for the estimation of the registration functions than the non-spatial alternative. Moreover, it is found that the driving time between locations represents the spatial dependency in the Covid-19 data better than geographical distance. However, since driving time is non-Euclidean, the traditional spatial frameworks break down; to solve this, a methodology inspired by multidimensional scaling is developed to approximate the driving times by a Euclidean distance which enables the established theory to be applied. Finally, the resulting estimates of the registration/warping processes are analysed by taking functionals to understand the qualitatively observable earliness/lateness and sharpness/flatness of the Covid-19 waves quantitatively.
\end{abstract}
\noindent
{\it Keywords:}  spatial statistics, functional data, time series, phase variation
\vfill

\newpage

\section{Introduction}
\label{sec:intro}

On 12 March 2020, Sir Patrick Vallance (the Chief Scientific Advisor to the UK Government) claimed at the UK's daily Covid-19 press briefing:
\begin{quotation}
\noindent On the curve, we are maybe four weeks or so behind [Italy] in terms of the scale of the outbreak. \citep{Yates_2020}
\end{quotation}
Meanwhile, John Burn-Murdoch of the Financial Times wrote on 11 March:
\begin{quotation}
\noindent Case numbers since outbreaks began in several countries have tracked $\sim 33\%$ daily rise. This is as true for UK, France, Germany as Italy; the latter is simply further down the path. \citep{Murdoch_2020}
\end{quotation}
During the Covid-19 pandemic, there was an explosion in the public discourse of the masses of data put out, but what do these above claims mean from a statistical standpoint? How can the available data be used to justify (or refute) these claims? Considering the Covid-19 waves in the UK and Italy as functions of time, the above statements are examples of descriptions of functional registration: the alignment of functional data by a registration process, which `corrects' for an assumed warping process causing the non-alignment of curves in measured time. In this case, the aim is to elicit this phase variation. In fact, John Burn-Murdoch in producing the FT's initial country comparison charts for Covid-19 translated time from calendar date to days since 100 confirmed cases, a form of time warping for interpretability.

\begin{figure}
\centering
\includegraphics[width=0.7\textwidth]{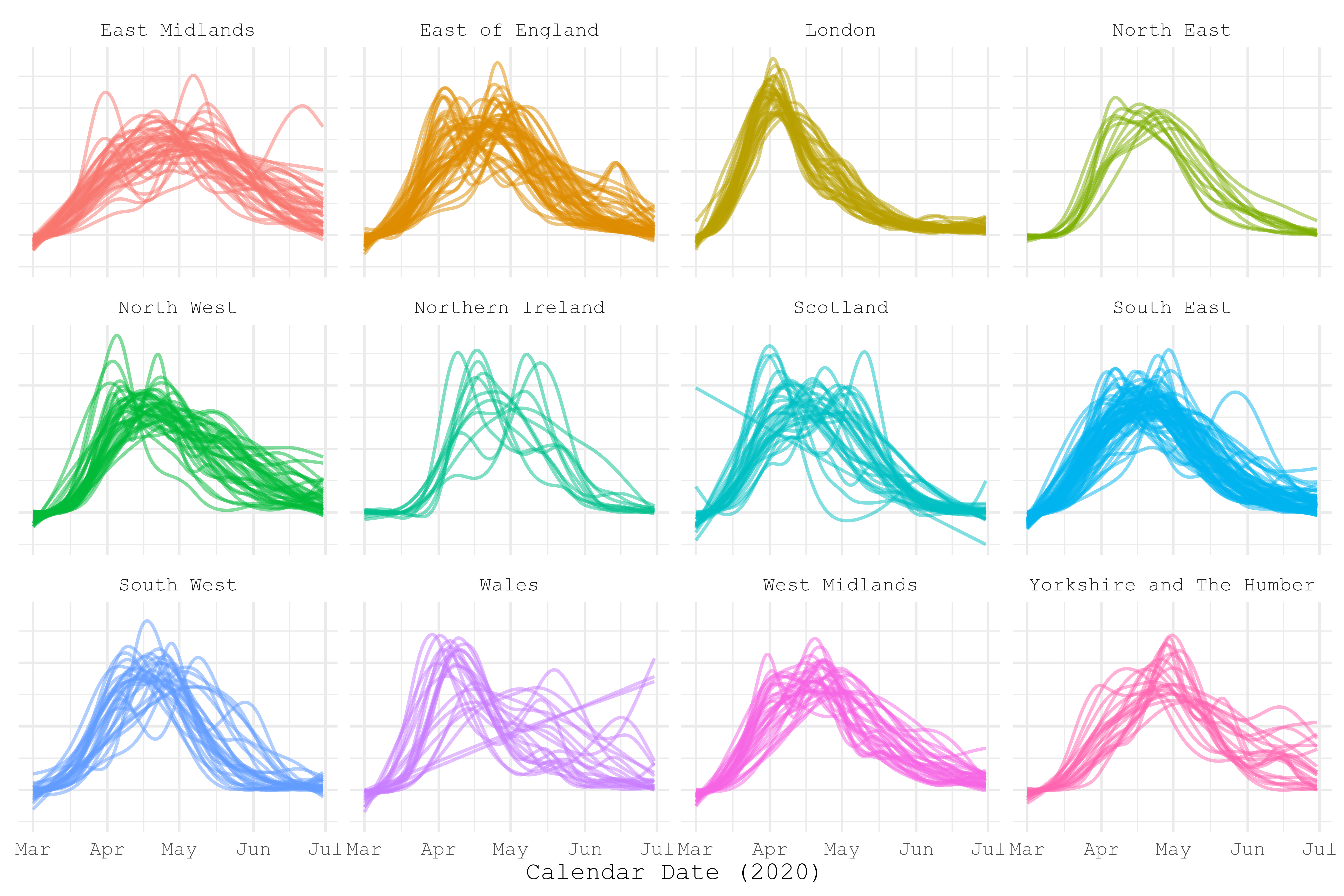}
\caption{Curves of SARS-CoV-2 incidence in the UK from 1 March to 30 June 2020, disaggregated by LTLA. The curves have been smoothed (cubic splines with REML for parameter selection) and normalised (by $L^2$ norm). The facets are the nine regions of England, Wales, Scotland and Northern Ireland. For a geographic reference, see Figure \ref{fig:ltla.locations}.}
\label{fig:covid.waves}
\end{figure}

Inspired by these statements, this paper seeks to understand how to estimate and interpret phase variation across geography, embracing a general diffeomorphic warping of a standard `central' Covid-19 wave, although these ideas could apply to any spatially indexed functional data. Functionals of these warpings may then be taken to understand the relataive displacement of the curve earlier or later, but also for example the relative sharpness or flatness of the curves. Figure \ref{fig:covid.waves} provides the example of the first wave of recorded SARS-CoV-2 incidence in the UK (by specimen date of a positive test result), disaggregated by lower-tier local authority (LTLA), a unit of geography in the UK, and normalised for visual comparison of the shapes, provided by the UK's Office for National Statistics (ONS). In this case, the shape of the curve is predominantly a single peak, with the majority of the variation represented by location and sharpness of this peak. This can most clearly be seen comparing the East Midlands to London, the former having a later and flatter peak, the latter having an earlier and sharper peak.

From a methodological standpoint, this paper investigates the problem of functional registration: Given observations of random functions assumed to be from a temporal domain $[0,1]\rightarrow\mathbb{R}$, there are sought registration processes that warp time $[0,1]\rightarrow [0,1]$  from the locally observed clocks (calendar date) to a latent global clock, under which the features of the curves are aligned (a sharp rise then steady fall in cases). Typically, this is justified either as removal of noise in the temporal observations, or simply as a facilitation of more parsimonious and comprehensible analyses of the amplitude variation. However, here the phase variation itself is the object of statistical interest, and is explicitly modelled and analysed; it is how early/late and sharp/flat the waves were that is of interest, not the height of the wave. Functional registration is a well studied problem with myriad solutions in the iid case, but here we are required to assume a spatial model, contradicting the assumption of independence. This not only complicates the theoretical backing of the existing methodologies, but also provides an opportunity to determine more efficient spatially-aware statistical procedures.

In the non-spatial case, a wide gamut of approaches have been suggested to solve this registration problem. To list but a handful, there is: landmark registration, where particular features of interest such as turning points and zeros of the functions are aligned and that in between is linearly warped \citep{kneip1992statistical}; template registration, where the curves are warped to closely align to a particular template, such as the current pointwise mean, and then one iterates to convergence \citep{ramsay1998curve}; quantile registration, where the curves are registered such that the proportion of the area under the curves are equal at each time point \citep{chretien2021quantile}; principal components based registration, where the registration process is constructed to maximise the parsimony of the subsequent functional principal component analysis \citep{kneip2008combining}; semi-parametric registration, where the form of the phase variation is deemed to be known up to a finite set of parameters, which are then estimated by likelihood methods \citep{hardle1990semiparametric}; registration by pairwise curve synchronisation, where pairs of curves are warped together and an average of these pairwise warpings is taken to estimate warping to a global clock \citep{tang2008pairwise}; registration under Fisher--Rao geometry, where a distance invariant to diffeomorphic time-warping is used to produce a Karcher mean function for the curves understood modulo time-warping, which is then used as a template for registration \citep{srivastava2011registration}; and local variation registration, where the distances swept in the abscissae are aligned at each time point \citep{chakraborty2021functional}. Any of these methods could be use to understand the phase variation in the first Covid-19 wave across geography, but it is the latter of these approaches that is developed in this paper, although, in Appendix A, developments of the penultimate and antepenultimate methodologies are also provided and compared.

At the heart of the differences between these approaches to functional registration is the solution of the question of identifiability: the same distribution of random functions may be obtained by the combination of various different pairs of distributions of phase and amplitude variation. This is explored in depth by \cite{chakraborty2021functional}. They demonstrate that a rank-one model for the amplitude variation with a mostly free model for the phase variation is sufficient for identifiability, but that almost any additional freedom in the amplitude model leads to issues of unidentifiability. All of the above papers must in some way overcome this issue, whether explicitly or not. The potential extrinsic rationales for imposing restrictions on the models of phase and amplitude variation are varied, which accounts to some degree for the wide array of approaches to functional registration and the continued popularity of the entire toolkit. For this paper, the identifiable regime given in \cite{chakraborty2021functional} is assumed---particularly appropriate for cases when there is no extrinsic rationale for the restriction of the phase variation model. For the first wave of the Covid-19 pandemic, any substantive restrictions on the phase variation (beyond perhaps some regularisation) seems difficult to justify; meanwhile, a unimodal amplitude model---and therefore rank-one amplitude variation---appears reasonable up until the start of the second wave.

\begin{figure}
\centering
\includegraphics[width=0.7\textwidth]{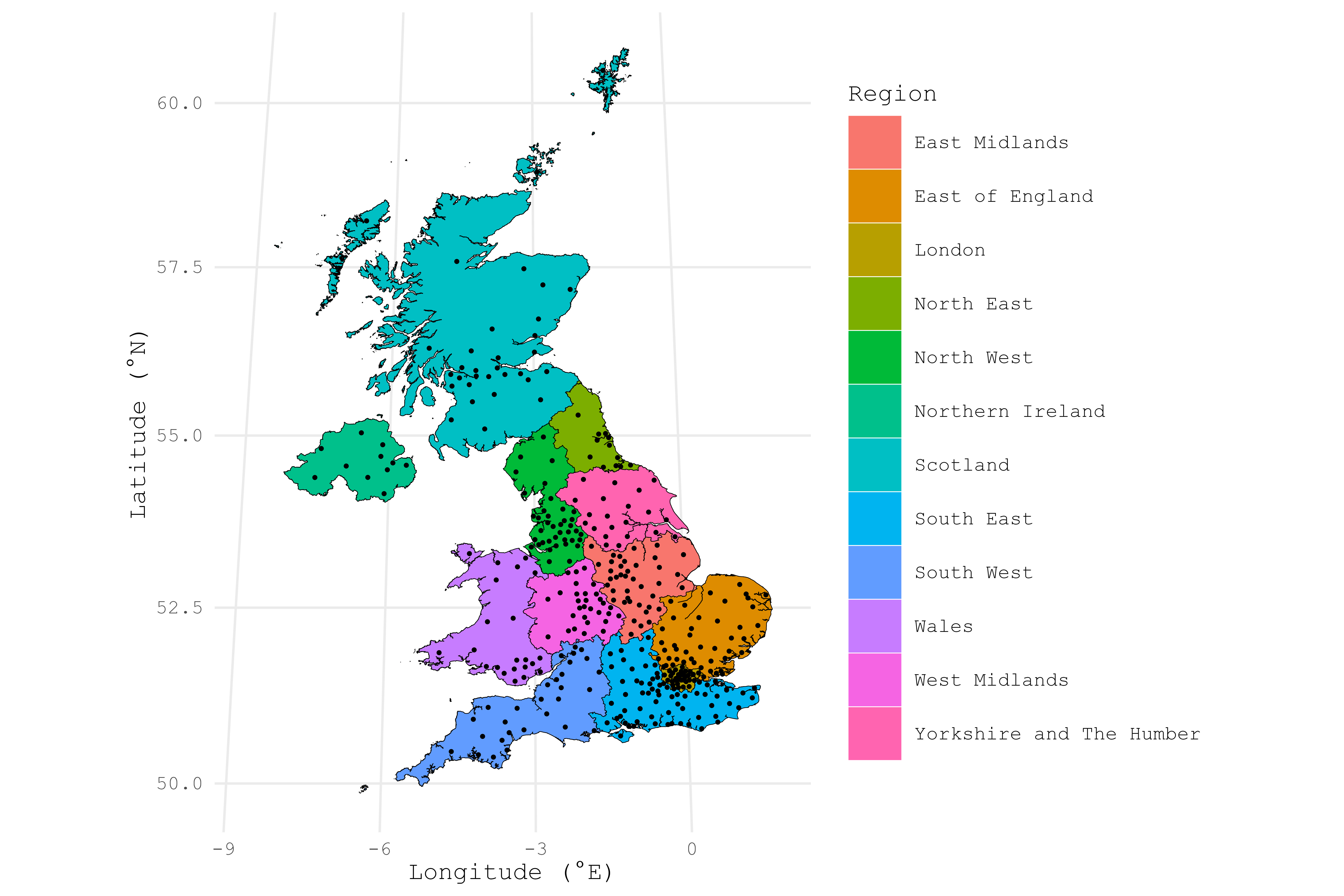}
\caption{Centroids of the LTLAs of the UK. Coloured are the nine regions of England, Wales, Scotland and Northern Ireland. Note the much greater density of LTLAs in London when compared to the more rural areas of the UK, such as Scotland.}
\label{fig:ltla.locations}
\end{figure}

The described problem of disentangling phase and amplitude variation is complicated further here by the explicitly spatial nature of the observations. This manifests as (typically positive) correlation between observations, depending on the spatial distance between them. If the Covid-19 wave in Hackney (London) is particularly early, its likely the wave was also early in Islington next door; meanwhile, it's unlikely that it has anywhere near as much to do with the wave in Ceredigion on the west coast of Wales. Thus, the assumption of independence of the observations must be rejected, leaving many of the existing procedures for registration unjustified. Moreover, there may be methodologies that exploit the spatial structure and therefore produce more efficient estimates of the warping process. This can be understood from an information-theoretic point-of-view: knowledge of one observation diminishes the marginal information gain from nearby observations; hence, an optimal methodology should increase the weight of observations at sparsely observed regions and decrease the weight of observations at densely observed regions. In the case of the Covid-19 data, the LTLAs at which the incidence is measured vary greatly in density; Figure \ref{fig:ltla.locations} provides a map of the centroids of LTLAs in the United Kingdom. An overview of geospatial statistical problems can be found in \cite{cressie2015statistics} or \cite{gelfand2010handbook}.

Similarly to the identifiability problem of functional registration where one must disentangle phase and amplitude variation by imposing some assumptions, in spatial statistics there is a problem of identifying between a mean structure varying across space and simply a correlated error structure. Here, this will be solved by assuming a level of stationarity on the data. This involves assuming a spatially independent mean structure and a covariance structure that depends solely on the distance between observations. This allows for the identifying of what is mean structure and what is covariance structure; otherwise, one would need an intrinsic reason to believe in the impact of some covariates for the mean structure or repeated independent observations over space.

The question remains though as to the choice of the distance metric $d$, through which the covariance structure depends on the observation locations. The typical choice is the Euclidean distance: it is both easy to compute and associated with a wide array of parametric families of covariograms which are positive definite. However, to understand Covid-19 dynamics, it may be more reasonable to think of other distance metrics, such as travel time, which might more accurately approximate the covariance structure. To use such a metric though one must establish the validity of the parametric family of covariograms. This is solved here by approximating this travel time metric by the Euclidean metric in a high dimensional space (see Section \ref{s:DistanceMetric}).

In the literature, there is one existing approach from \cite{guo2023spatially} to register curves in a spatially-aware way. They take into account space via a spatially penalised registration, in effect a form of template registration where the templates are locally generated. While this method is useful, it however, not only requires the introduction of a spatial smoothing parameter, but the algorithm is also not proven to produce consistent estimators of the registration process---nor even converge. The approach below is more rigorously justified, being a modification of an existing, theoretically established framework.

Even though Covid-19 is now controlled through vaccination and acquired immunity through exposure, the questions of timing are still of importance. In any future pandemic (or more localised infectious disease outbreak), the notion of timing is critical. The ability to address questions around relative timings of different geographical locations, and their (in)dependence is crucial for establishing the right policy responses, and drives the development of the methodology below.

\section{The Statistical Model}

The observable in the model, representing the rate of Covid-19 transmission, is a random functional field, $Y:\Omega\times\mathcal{D}\rightarrow L^2[0,1]$, defined on a probability space $(\Omega, \mathcal{F}, \mathbb{P})$. Here, $\mathcal{D}$ is the spatial domain (clasically $\mathbb{R}^2$ or $\mathbb{R}^3$ endowed with the Euclidean distance, but here an arbitrary metric space is accepted), and $[0,1]$ is the temporal domain. In all future notation, dependence on $\omega\in\Omega$ will be suppressed, and the distinction between the random variables and their realised counterparts will generally be in the traditional cases (for example $y=Y(\omega,\cdot)$). This random functional field is observed at spatial locations $(s_i)_{i=1}^n$: the lower tier local authorities (LTLAs). For ease of notation, let $Y_i:[0,1]\rightarrow\mathbb{R}$ be defined as $Y_i:=Y(s_i)$. These functions are observed at temporal locations $(t_{ij})_{j=1}^{m_i}$; let $Y_{ij}\in\mathbb{R}$ be defined as $Y_{ij}:=Y_i(t_{ij})$ (in the measurement error-free case). These are daily observed incidence of SARS-CoV-2 for each LTLA from 1 March to 30 June 2020. Note that temporal observation locations may in general depend on spatial location, but in the case of the Covid-19 data it does not. For consistency, the indices $i$, $k$ and $\ell$ will refer to spatial locations and $j$ to temporal locations; spatial locations in general may be referred to as $s$, measured temporal locations as $t$, and latent global clock (from which time is warped) temporal locations as $\tau$.

This random functional field $Y$ is decomposed into two components:
\begin{equation}
Y(s) = X(s) \circ H^{-1}(s),\quad s\in\mathcal{D}.
\end{equation}
Here, $X:\mathcal{D}\rightarrow L^2[0,1]$ is a random functional field representing the amplitude variation in a latent global clock under which features of interest are aligned. On the other hand, $H:\mathcal{D}\rightarrow \mathcal{W}$ is a random functional field representing the phase variation: it provides the transformations that warp the latent global clock under which features are aligned into the observed local clocks. Here, $\mathcal{W}$ represents the set of all valid warping functions, explained below. Similar to $Y$, let $X_i:=X(s_i)$ and $H_i:=H(s_i)$.

It may be assumed that $Y$ is in fact observed with measurement error:
\begin{equation}
Y_{ij} = X_i \circ H_i^{-1} (t_j) + \epsilon_{ij},\quad j\in [m_i],\ i\in[n],
\end{equation}
where the $\epsilon_{ij}$ are mean-zero random variables representing the measurement error. A preprocessing step of smoothing is performed to estimate $Y_i$ directly.

Now the individual components, $X$ and $H$, are understood. First, the amplitude variation is assumed to be rank-one. That is, it is of the following form:
\begin{equation}
X(s) = \Xi(s)\mu,\quad s\in\mathcal{D},
\end{equation}
where $\Xi:\mathcal{D}\rightarrow\mathbb{R}^+$ is a random scalar field with unit mean and $\mu:=\mathbb{E}X(s)\in L^2[0,1]$ is a non-stochastic mean function. Let $\Xi_i=\Xi(s_i)$. Note there is an implicit assumption that $\Xi$ has no drift, so that $X$ has no drift (that is, has a constant mean in space). While this rank-one model may seem restrictive, it is conceivable in many settings there may be extrinsic rationale for belief in this model, which is often not the case for restrictions on the model for $H$; by enforcing this restriction to the model for $X$, by \cite{chakraborty2021functional} there may be a very flexible model for $H$. As motivation, the first Covid-19 wave in England may be thought of as a unimodal wave---of rank-one variation---warped into different shapes and positions, but retaining its unimodality.

Compared to such a restrictive amplitude model, only two assumptions are made in the model for $H$. First, for each spatial location $s\in\mathcal{D}$, $H(s)$ is an increasing diffeomorphism from $[0,1]$ to $[0,1]$. This means there are no time-reversal, plateaus or jumps, and the start and end times of the local clocks are already aligned. These assumptions may often be extrinsically justified. This provides the set of valid warping functions, $\mathcal{W}$, which indeed is a group with respect to functional composition. The second assumption made is that $\mathbb{E}H(s)=\mathrm{id}$ for all $s\in\mathcal{D}$: there is in expectation no warping from the latent global clock. While making the inference on $H$ more interpretable, this is in fact not a restriction on the model for $Y$, but rather merely a requirement for identifiability. Note again there is an assumption of no drift in $H$.

Finally, it is to be assumed that the spatial correlation structure is a function of distance between observation locations only. If this were a Euclidean distance, this is stationarity and isotropy. Indeed, stationarity and isotropy lead to a particularly nice theory of correlation structures, and valid covariogram families, satisfying positive definiteness. Thus, although it is assumed below that a non-Euclidean distance may be the `true' parameter on which the covariance structure depends, a Euclidean approximation is utilised to allow the use of this theory. In particular, the Matérn model (or a subfamily of the Matérn model) will be used \citep{matern1960spatial} which is of the form:
\begin{equation}
C(d;\sigma^2,\nu,\rho) = \iota\mathbb{I}(d=0) + \sigma^2 \frac{2^{1-\nu}}{\Gamma(\nu)}\left(\sqrt{2\nu}\frac{d}{\rho}\right)^\nu K_\nu\left(\sqrt{2\nu}\frac{d}{\rho}\right),
\end{equation}
where $\Gamma$ is the gamma function and $K_\nu$ is the modified Bessel function of the second kind. The parameters are all positive, where $\iota$ is a nugget representing the inherent variation observing at one point, $\sigma^2$ is the semisill representing an overall measure of spatial variation, $\nu$ is a shape parameter which corresponds to smoothness and $\rho$ is a range parameter which corresponds to how quickly observations become less correlated at increasing distances. In Covid-19 terms, $\rho$ represents a typical distance determining how far apart the spatial dynamics extend, while $\nu$ represents the volatility of the observations between neighbouring authorities. This Matérn family is valid for a Euclidean distance in any number of dimensions. For more information of isotropic covariogram families, see for example  \cite{gelfand2010handbook}.

\section{The Distance Methodology}\label{s:DistanceMetric}

The choice of distance in the spatial methodology is crucial, since the assumption that the variance between observations at two locations is a function of this distance is quite restrictive. It is therefore important to ensure the distance used indeed does capture the spatial covariance structure to the greatest degree possible. However, while non-Euclidean distance metrics are often used in applications, there are substantial mathematical issues with their use, which are often not overcome. In particular, the particular form of the Euclidean distance as a function of a displacement vector with the spatial observations living in a vector space allows the justification of various parametric covariogram families which satisfy the condition of positive definiteness. This arises from an assumption of stationarity and isotropy. Often the validity of a covariogram family depends on the dimensionality of the data, but the Matérn family, for example, is valid in all dimensions. On the other hand, the use of a non-Euclidean distance metric with such a family in general leads to estimates of covariance matrices with negative eigenvalues.

However, in practice it is often more reasonable to assume that the covariogram is a function of a non-Euclidean distance metric, rather than some alternative Euclidean distance (if some Euclidean embedding is even apparent). In the case of Covid-19, for example, it may be more reasonable to assume that the epidemiological dynamics spread on the scale of human travel rather than kilometres, which may be approximated by driving time between locations. Indeed, as will be found in the application section below, the use of travel time does indeed elucidate the spatial structure in a way kilometer distance between locations does not. It is necessary therefore to either justify the use of a particular covariogram family for this particular non-Euclidean distance metric, or to approximate the non-Euclidean distance by a Euclidean distance metric under some embedding in a Euclidean vector space. In this paper, the latter course is chosen due to the intractability of the choice of non-Euclidean distance.

Suppose $D$ is an $n\times n$ distance matrix with corresponding centred inner-product matrix
\begin{equation}
B := -\frac{1}{2}H(D\circ D) H,
\end{equation}
where $H$ is the centring matrix $I_n-\frac{1}{n}1_n1_n^\top$ and $\circ$ is the Hadamard (element-wise) product. Note that for an arbitrary rank-$p$ positive semi-definite matrix $B^*\in\mathbb{R}^{n\times n}$ with spectral decomposition $B^*=\sum_{i=1}^p u_i\lambda_iu_i^\top$ for $u_i\in\mathbb{R}^n$ and $\lambda_i>0$, one can define $n$ locations in $\mathbb{R}^p$:
\begin{equation}
s_i:=\left(\lambda_1(u_1)_i,\ldots,\lambda_p(u_p)_i\right)^\top,\quad i\in[n],
\end{equation}
with corresponding Euclidean distance matrix $D^*$ such that $B^*=-\frac{1}{2}H(D^*\circ D^*) H$. Indeed, the centred inner-product matrix is positive semi-definite if and only if $D$ is a Euclidean distance matrix. This suggests approximating $B$ so that it's positive definite, rather than the more intractable task of the approximation of $D$ so that it's Euclidean. The methodology therefore works out as in Algorithm \ref{alg:dist}, similar to methods in multi-dimensional scaling.
\begin{algorithm}
\label{alg:dist}
\caption{To approximate a (non-Euclidean) distance matrix $D\in\mathbb{R}^{n\times n}$ by a Euclidean distance matrix $D^*\in\mathbb{R}^{n\times n}$.}
Set $B:= -\frac{1}{2}H(D\circ D) H$\;
Compute the spectral decomposition $B =: \sum_{i=1}^n u_i\lambda_i u_i^\top$, where $\lambda_1\geq\cdots\geq\lambda_n$\;
Choose $p\leq \rank B$\;
Set $s_i^* = (\lambda_1(u_1)_i,\ldots,\lambda_p(u_p)_i)^\top$ for $i\in[n]$\;
Set $D^*_{ik}:= ||s_i^* - s_k^*||$ for $i,k\in[n]$\;
\end{algorithm}

All that remains is then the selection of $p$. There are four criteria for the choice of $p$ suggested here:
\begin{enumerate}
\item The distance-based approach: choose the value of $p$ such that a matrix distance between $D$ and $D^*$ is minimised. This has the benefit of maximising the interpretability of the new distance matrix $D^*$ as an approxmation of an extrinsically-justified choice of non-Euclidean distance.
\item The data-based approach: choose the value of $p$ such that the sill-to-nugget ratio of the resulting variogram estimate is maximised. This criteria suggests that the chosen approximation maximises under the methodology the spatial explanation of the randomness in the data compared to the inherent variation.
\item If a particular variogram model is to be used, one must be careful to restrict $p$ to be no greater than the maximum dimensionality in which the variogram model is valid. For example, if the spherical correlation function is used (for example to enable computationally efficient estimation), one is restricted to $p\leq 3$.
\item Choose $p=2$ or $p=3$ so that the data can be visualised easily. (Note that spatial methodologies may be applied using a higher dimension, and graphically displayed by this methodology in a lower dimension.)
\end{enumerate}

Once this Euclidean distance matrix $D^*$ is determined, the standard spatial statistics methodologies can be applied; for this paper on Covid-19 spatiotemporal dynamics this is a spatial registration, but it could more generally be utilised as well.

\section{The Registration Methodology}

This section develops a spatially aware functional registration methodology, which can be used with any Euclidean distance metric, be it an approximation via Algorithm \ref{alg:dist} or a real Euclidean distance. It is broken into three sections: the estimation of the full curves, the idea of the registration methodology, and its spatial adaptation. For a succinct overview of the method, see Algorithm \ref{alg:reg} at the end of the section.

\subsection{From Discrete Observations to Full Curves}

Before performing functional data analysis techniques, the first step is to estimate the full functions $Y_i\equiv Y(s_i):[0,1]\rightarrow\mathbb{R}$ for $i\in[n]$ given the observations $Y_{ij}\equiv Y_i(t_{ij})+\epsilon_{ij}\in\mathbb{R}$ for $j\in[m_i]$. There are myriad standard approaches to this problem. Given there is assumed measurement error---which may well be autocorrelated---a smoothing method is chosen: cubic smoothing splines with smoothing selected by restricted maximum likelihood. In particular, this involves setting the smoothing splines:
\begin{equation}
\hat{Y}_i^{(\alpha)}:=\underset{\tilde{y}\in C^2[0,1]}{\argmin}\left[ \sum\limits_{j=1}^{m_i} \left(Y_{ij} - \tilde{y}(t_{ij})\right)^2 + \alpha\int_0^1 \left( \mathrm{D}^2\tilde{y}(t)\right)^2\mathrm{d}t \right],
\end{equation}
where $\alpha\in\mathbb{R}^+_0$ is an appropriate smoothing constant chosen by restricted maximum likelihood. This method of smoothing parameter selection was chosen due to the probable heteroscedastic and autocorrelated noise; for an overview of different tuning methods, see \cite{berry2021cross}. The estimated full curves are then:
\begin{equation}
\hat{Y}_i := \hat{Y}_i^{\left(\hat{\alpha}_i\right)}.
\end{equation}
Note that due to the smoothing these are in fact estimates of the error-free $Y_i$, and so the following analysis can maintain focus on disentangling phase and amplitude variation.

Also required are estimates of the derivatives $\mathrm{D}Y_i$, which are obtained by quintic smoothing splines: the solution to the following minimisation problem:
\begin{equation}
\widehat{\mathrm{D}Y}^{\left(\beta\right)}_i:=\mathrm{D}\left(\underset{\tilde{y}\in C^4[0,1]}{\argmin}\left[ \sum\limits_{j=1}^m \left(Y_{ij} - \tilde{y}(t_{ij})\right)^2 + \beta\int_0^1 \left( \mathrm{D}^3\tilde{y}(t)\right)^2\mathrm{d}t \right]\right),
\end{equation}
where $\beta\in\mathbb{R}^+_0$ is an appropriate smoothing constant also chosen by restricted maximum likelihood. The estimated derivatives are then:
\begin{equation}
\widehat{\mathrm{D}Y}_i:=\widehat{\mathrm{D}Y}^{\left(\hat{\beta}_i\right)}_i.
\end{equation}

For a discussion of the use of different orders of smoothing splines for the estimation of functions and their derivatives, see \cite{cao2012estimating}, for example. These methodologies rely on the assumption of densely observed data, such that each $Y_i$ is well-described by the observations $Y_{ij}$. In the sparse case, there are other methodologies available such as PACE, to borrow information across the curves. This added complication is not considered here as this is not required for the Covid-19 analysis. See \cite{liu2009estimating} for a more detailed look at this problem.

\subsection{Local Variation Analysis: The Idea}

The approach to registration of local variation analysis was introduced by \cite{chakraborty2021functional}, who also demonstrated convergence properties. The idea is based on the consideration of the local variation distribution of a curve $f$:
    \begin{equation}
        \Lambda_f(t) := \frac{\int_0^t |\mathrm{D}f(\tilde{t})|\mathrm{d}\tilde{t}}{\int_0^1 |\mathrm{D}f(\tilde{t})|\mathrm{d}\tilde{t}},\quad f\in C^1[0,1],\quad t\in [0,1],
    \end{equation}
    where $C^1[0,1]$ represent continuously differentiable functions on the temporal domain $[0,1]$. This distribution encapsulates abscissae variation with respect to the ordinate, that is how much the incidence varies from day to day. The key feature of this function is that---as well as being scale-invariant:
    \begin{equation}
        \Lambda_{f\circ \theta^{-1}} = \Lambda_f  \circ \theta^{-1},\quad f\in C^1[0,1],\quad \theta\in\mathcal{W}.
    \end{equation}
    Thus:
    \begin{equation}
        \mathbb{E} \Lambda_{X_i\circ H_i^{-1}}^{-1} = \mathbb{E}\Lambda_{X_i}^{-1} = \Lambda_{\mu}^{-1}.
    \end{equation}
    This means that $\Lambda_\mu$ can be directly estimated from the $X_i\circ H_i^{-1}$ (which can be estimated by smoothing splines), whence one can obtain:
    \begin{equation}
        H_i = \Lambda_{X_i\circ H_i^{-1}}^{-1} \circ \Lambda_\mu.
    \end{equation}
    Finally, $X_i=Y_i\circ H_i$.

    The spatial modification to this methodology comes in the estimation of the $\Lambda_\mu^{-1}$ from the $\Lambda_{X_i\circ H_i^{-1}}^{-1}$; in the non-spatial case, this is done by a simple mean, but in the spatial case a weighted mean will be utilised, with weights obtained from a variogram model. This is where densely observed areas such as London will be downweighted since they provide marginally less information regarding the central local variation distribution than the sparsely observed areas.

\subsection{Estimation of a Spatial Mean}

The key estimation in the above is that of the mean of a random functional field. The best linear unbiased estimator of this mean is sought; that is, one must optimise the estimator $\sum_{i=1}^n w_iZ_i$ of $\mathbb{E}Z$ (independent of $i$) over all $(w_i)_{i=1}^n$ such that $\sum_{i=1}^n w_i=1$ (i.e. the estimator is unbiased), in terms of mean squared error. This is a problem considered by \cite{gromenko2013nonparametric}, and by the method of Lagrange multipliers one obtains the following optimal weights:
\begin{equation}
w = \frac{C^{-1}1_n}{1_n^\top C^{-1}1_n},\quad C_{ik} = \mathbb{E}\langle Z_i - \mathbb{E}Z, X_k - \mathbb{E}Z\rangle,
\end{equation}
where $w$ is the length-$n$ column vector of weights, $C$ is the $n\times n$ matrix of covariances, and $1_n$ is the length-$n$ column vector of ones.

Given the stationarity of the model with respect to the approximating Euclidean space, one hase that $C_{ik}=C(d_{ik})$ where $d_{ik}=d(s_i,s_k)$ for the approximating Euclidean distance $d$. This function $C:\mathbb{R}_0^+\rightarrow\mathbb{R}$, called the covariogram, arises as:
\begin{equation}
2\gamma(d_{ik}):=\mathbb{E}||Z_i-Z_k||^2 = 2\gamma(\infty)-2C(d_{ik}),\quad i,k\in[n].
\end{equation}
Thus, one needs only use the data $(d_{ik},||Z_i-Z_k||^2)_{i\neq k}$ to estimate the variogram $\gamma:\mathbb{R}_0^+\rightarrow\mathbb{R}_0^+$ and thereby the covariogram $C$.

\subsection{The Spatial Methodology}

Now the idea of how local variation analysis works has been introduced, the exact details of how it is implemented are described below. Moreover, the spatial estimation of the $\Lambda_\mu^{-1}$ is provided in detail.

First estimated are the $\Lambda_i:=\Lambda_{Y_i-\epsilon_i}$:
    \begin{equation}
        \hat{\Lambda}_i (t) := \frac{\int_0^t \left|\widehat{\mathrm{D}Y}_i(\tilde{t})\right|\mathrm{d}{\tilde{t}}}{\int_0^1 \left|\widehat{\mathrm{D}Y}_i(\tilde{t})\right|\mathrm{d}\tilde{t}}.
    \end{equation}
    Note once again the quintic smoothing splines are used for derivative estimation. The inverses $\hat{\Lambda}^{-1}_i$ are then computed.

    A weighted mean of these is then used to estimate $\Lambda_\mu^{-1}$:
    \begin{equation}
        \left(\hat{\Lambda}_\mu^{(\tilde{w})}\right)^{-1} := \sum\limits_{i=1}^n \tilde{w}_i \hat{\Lambda}_i^{-1}.
    \end{equation}
    The optimal choice is $\tilde{w}$ is defined to be:
    \begin{equation}
        w := \underset{\substack{\tilde{w}\in\mathbb{R}^{n}: \\ \sum\limits_{i=1}^n\tilde{w}_i=1}}{\argmin} \mathbb{E}\left|\left|
\left(\hat{\Lambda}_\mu^{(\tilde{w})}\right)^{-1} - \Lambda_\mu^{-1} \right|\right|_2^2 = \frac{C^{-1}1_n}{1_n^\top C^{-1}1_n},
    \end{equation}
    where $C\in\mathbb{R}^{n\times n}$ is the matrix defined by its entries:
    \begin{equation}
        C_{ik}:=\mathbb{E}\left\langle \Lambda_i^{-1} - \Lambda_\mu^{-1}, \Lambda_k^{-1} - \Lambda_\mu^{-1} \right\rangle.
    \end{equation}
    Extending these covariances $C_{ik}$ to the covariance function $C:\mathcal{D}\times\mathcal{D}\rightarrow\mathbb{R}$, one can utilise the stationarity of the observations to obtain $C(s,r)=C(d(s,r))$ for any $s,r\in\mathcal{D}$. Moreover, defining the variogram:
    \begin{align}
        2\gamma(d(s,r)) &:=\mathbb{E} \left|\left|\Lambda_{Y(s)}^{-1} - \Lambda_{Y(r)}^{-1}\right|\right|^2\\ &= 2\mathbb{E}\left|\left|\Lambda_{Y(s)}^{-1} - \Lambda_\mu^{-1} \right|\right|^2 - 2\mathbb{E}\left\langle\Lambda_{Y(s)}^{-1} - \Lambda_\mu^{-1}, \Lambda_{Y(r)}^{-1} - \Lambda_\mu^{-1} \right\rangle\\ &= 2\gamma(\infty) - 2C(d(s,r)).
    \end{align}

    This $\gamma$ is estimated by considering the data $\left(d_{ik}, ||\hat{\Lambda}_i^{-1}-\hat{\Lambda}_k^{-1}||^2\right)_{i\neq k}$ and performing iteratively reweighted nonlinear least squares on the Matérn model:
    \begin{equation}
        \gamma(d) = \iota + \sigma^2 - \sigma^2\frac{2^{1-\nu}}{\Gamma(\nu)}\left(\sqrt{2\nu}\frac{d}{\rho}\right)^\nu K_\nu\left(\sqrt{2\nu}\frac{d}{\rho}\right).
    \end{equation}
The target weights in the iteratively reweighted least squares are proportional to $(\gamma(d_{ik}))^2$ as demonstrated by Cressie, and so the iteration begins with uniform weights and updates in the obvious way.

    Once the estimates $\hat{\iota}$, $\hat{\sigma}^2$, $\hat{\nu}$ and $\hat{\rho}$ are obtained---and thereby $\hat{\gamma}$---one can estimate $\hat{C}_{ik}:=2(\hat{\sigma}^2 + \hat{\iota}) - 2\hat{\gamma}(d_{ik})$, and thereby $\hat{w}:= C^{-1}1_n\left/ 1_n^\top C^{-1}1_n\right.$, and thereby $\hat{\Lambda}_\mu := \hat{\Lambda}_\mu^{(\hat{w})}$.

Finally, one can obtain:
    \begin{equation}
        \hat{H}_i^{-1} := \hat{\Lambda}_\mu^{-1} \circ \hat{\Lambda}_i,
    \end{equation}
    whence one can also produce:
    \begin{equation}
        \hat{X}_i := \hat{Y}_i\circ \hat{H}_i.
    \end{equation}

\begin{algorithm}
\label{alg:reg}
\caption{To estimate the warpings $H_i^{-1}$ from the observations $Y_{ij}$ with pairwise distances $d_{ik}$.}
Fit quintic splines with REML smoothing to the $Y_{ij}$ and differentiate for $\widehat{\mathrm{D}Y}_i$\;
Compute $\hat{\Lambda}_i(t):=\left.\int_0^t \left| \widehat{\mathrm{D}Y}_i(\tilde{t}) \right|\mathrm{d}\tilde{t}\right/\int_0^1 \left| \widehat{\mathrm{D}Y}_i(\tilde{t}) \right|\mathrm{d}\tilde{t}$\;
Compute the variogram data $(d_{ik},\frac{1}{2}||\hat{\Lambda}_i^{-1}-\hat{\Lambda}_k^{-1}||^2)_{i\neq k}$\;
Fit the Matérn model by iteratively reweighted least squares, producing $\hat{\gamma}$\;
Estimate the covariances $\hat{C}_{ik}:=2\hat{\gamma}(\infty)-2\hat{\gamma}(d_{ik})$\;
Estimate the optimal weights $\hat{w}:=C^{-1}1_n\left/ 1_n^\top C^{-1}1_n\right.$\;
Compute $\hat{\Lambda}_\mu := \left(\sum_{i=1}^n \hat{w}_i \hat{\Lambda}_i^{-1}\right)^{-1}$\;
Estimate the warpings $\hat{H}_i^{-1}:=\hat{\Lambda}_\mu^{-1}\circ\hat{\Lambda}_i$\;
\end{algorithm}

\section{Simulations}\label{s:Simulations}

In order to evaluate the practical efficacy of the developed methodology, simulations of $Y$ were generated over a range of covariance structures and spatial observation locations, as described in the first subsection below. Both the non-spatial and spatial variants of the methodology were carried out, and the comparison of these is provided in the second subsection below. In particular, the average mean squared error,
\begin{equation}
    \frac{1}{n}\sum\limits_{i=1}^n\mathbb{E}\left|\left|\hat{H}_i^{-1} - H_i^{-1}\right|\right|_2^2,
\end{equation}
was estimated, demonstrating an improvement for the spatially-aware methodology. This justifies its use, leading to the application to the UK Covid-19 data in the next section.

\subsection{Data Generation}

\begin{figure}
\centering
        \centering
        \includegraphics[width=0.7\textwidth]{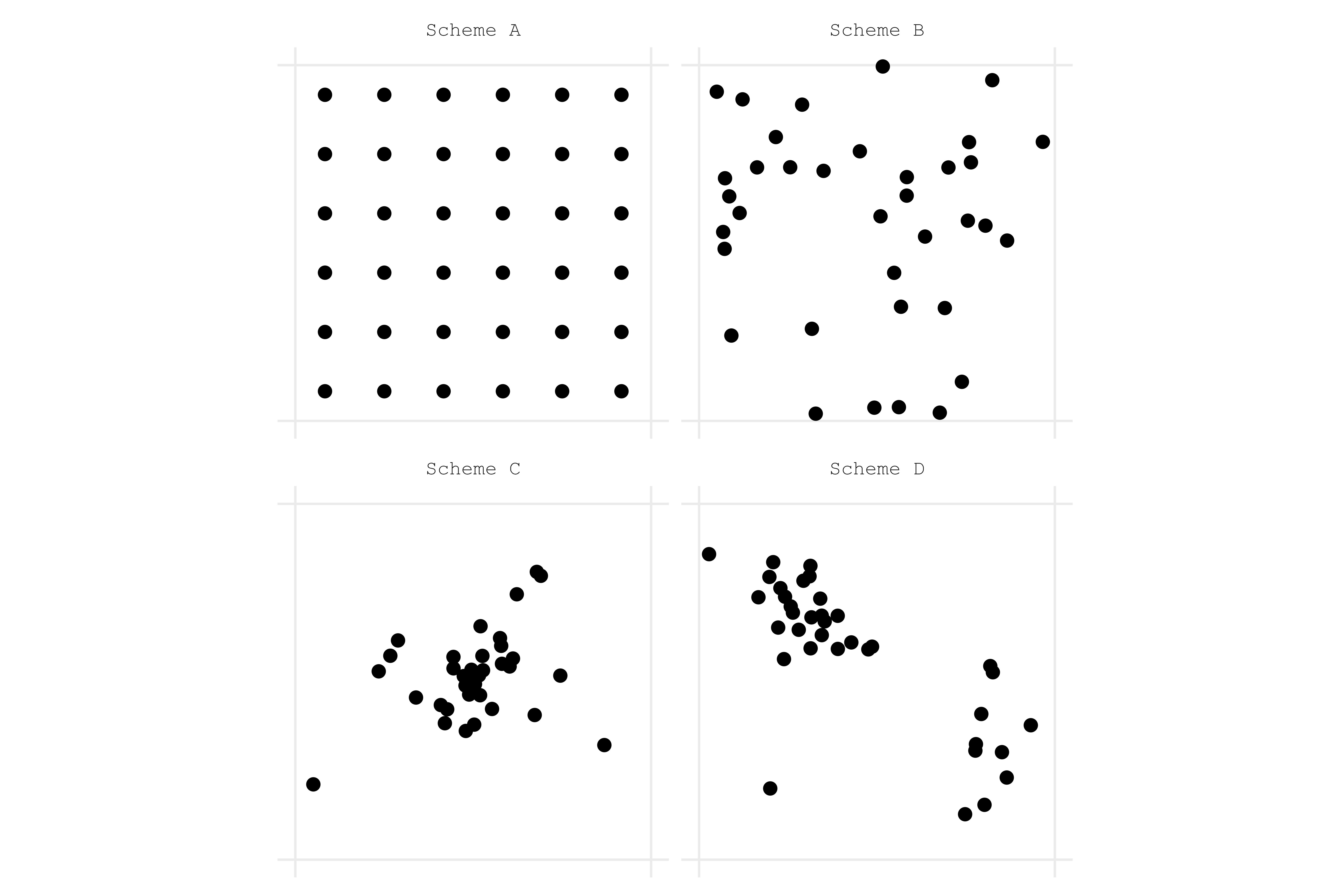}
        \caption{The simulated spatial observation locations in $\mathcal{D}=[0,1]^2$ (the box), for each of the four spatial observation schemata. Scheme A: square grid. Scheme B: uniformly at random. Scheme C: concentrated at centre. Scheme D: a mixture of bivariate normals.}
        \label{fig:schemata}
\end{figure}

Four schemata of spatial sampling were utilised, as plotted in Figure \ref{fig:schemata}. Each consists of 36 observation locations in $\mathcal{D}=[0,1]^2$. For details on their generation, see Appendix B.

These four schemata were chosen as representatives of the variety of potential real-world observation scenarios. For instance: scheme A may arise in image data, each location a pixel; scheme B may arise in point process data such as earthquakes; scheme C may arise data collected from a central location; and scheme D may arise in official data with each component of the mixed model a population hub. In principle, any of these types of spatial layouts could apply to COVID-19 data, depending on it collection protocol, but schemata C and D are of particular relevance to the data from the UK.

The temporal observation scheme was kept consistent as one hundred equispaced temporal observations, similar to the Covid-19 data which consists of 122 equispaced days.

The key variation in the covariance structures is in the range: the typical distance at which observations are $95\%$ uncorrelated. In particular, an exponential model was chosen of the following form:
\begin{equation}
    C(d) = 0.1\mathbb{I}(d=0) + \exp\left(-\frac{d}{\psi}\right),
\end{equation}
where $\psi$ ranged over $(0.03, 0.1,0.3,1.0)$, therefore ranges of $(0.09,0.29, 0.87, 2.90)$. (Note for reference that $d\in[0,\sqrt{2}]$ as the domain is the unit square.) The details as to how this covariance structure fed into the generation of the $h_i^{-1}$, see Appendix B.

The $x_i$ were chosen as:
\begin{equation}
    x_i(t) := \xi_i\sin(\pi t),
\end{equation}
where $\xi_i\overset{\mathrm{iid}}{\sim}\mathsf{Normal}(1, 0.04)$. Then the noise $\epsilon_{ij}\overset{\mathrm{iid}}{\sim} \mathsf{Normal}(0, 0.004)$ was added to produce:
\begin{equation}
    y_i = x_i\circ h_i^{-1} + \epsilon_i.
\end{equation}

\subsection{Results}

These data were simulated 5000 times---for each combination of observation scheme and covariance structure---and the local variation methodology was applied with and without the spatial correction. Some numerical instabilities led to the rejection of some simulations, particularly for the regularly spaced grid of Scheme A and for $\psi=0.03$, i.e. virtually independent observations; in these cases new simulations were generated to ensure that there were 5000 simulations for the comparisons. As can be seen in the results, there is convergence in mean squared error between the spatial and non-spatial variants as the true $\psi$ tended to zero (i.e. the methodologies produce equivalent results in the independent case).

\begin{figure}
\centering
\includegraphics[width=0.7\textwidth]{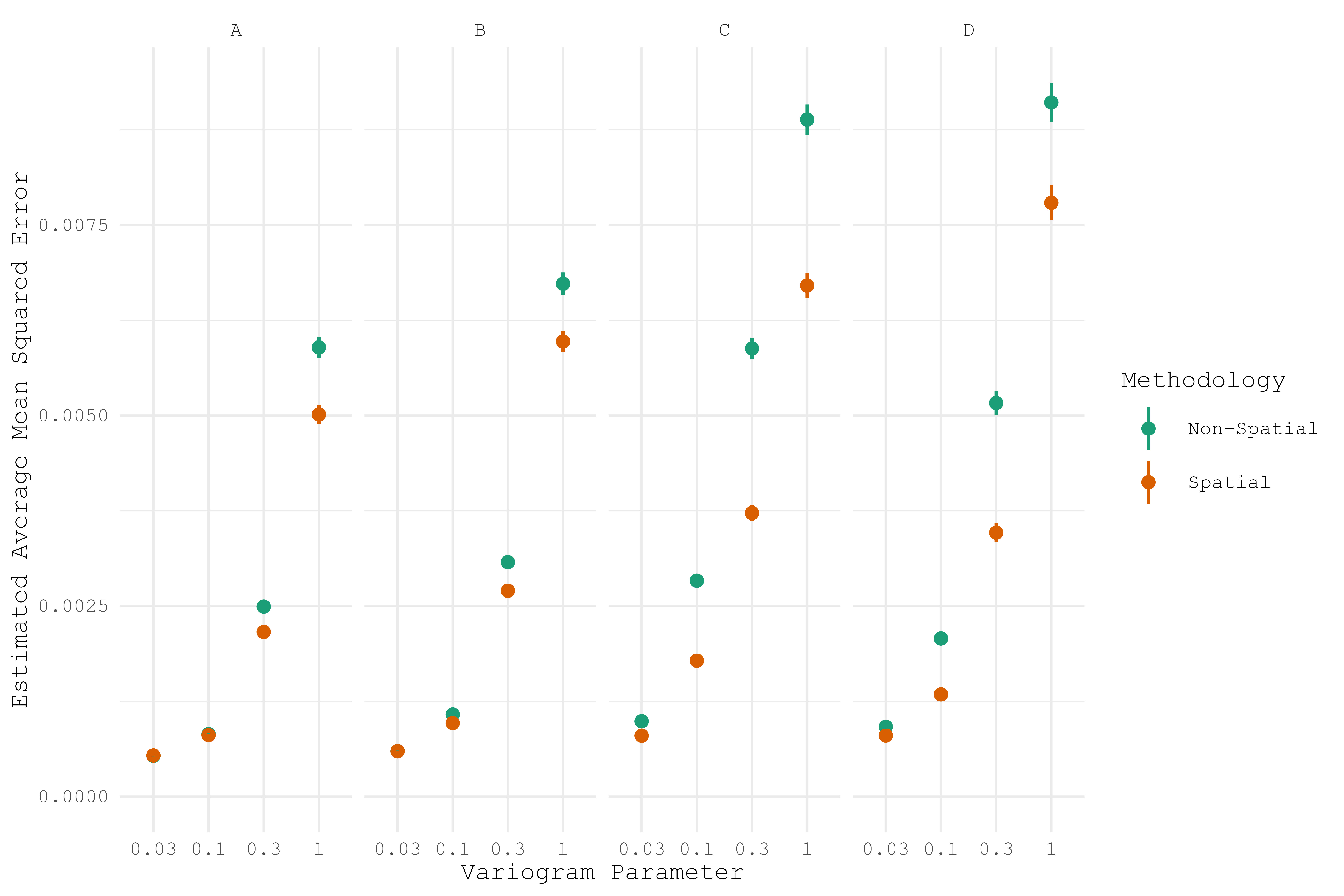}
\caption{Estimated average mean squared error in the estimation of the registration processes, for the four schemata and four variogram parameters for each of the spatial and non-spatial methodologies, with 95\% error bars.}
\label{fig:sim.out}
\end{figure}

The results of these simulations, in terms of the estiamted average mean squared error,
\begin{equation}
    \frac{1}{n}\sum\limits_{i=1}^n\mathbb{E}\left|\left|\hat{H}^{-1}_i - H^{-1}_i\right|\right|^2,
\end{equation}
are given in full in Appendix B, but are graphically plotted here with the 95\% confidence interval radius in Figure \ref{fig:sim.out}.

In general it can be seen the addition of a spatially-aware component to the methodology improves performance in estimating the $h_i^{-1}$. This is especially visible when $\psi$ is larger, implying a longer range of the covariance structure. This is not surprising, since then the effect of density of observation would have a greater effect on the marginal information obtained from each observation. It is also notable the greater the variation in density of observations, the greater the performance of the spatial methodology.

\section{Covid-19 Geographical Dynamics}

\subsection{The Curves}

There is considerable information around temporal disease dynamics in the daily SARS-CoV-2 infection trajectories by lower tier local authority (LTLA) in the UK, treated as a proxy for the Covid-19 wave curves in the UK for the period March--June 2020. These data come from the UK's Office for National Statistics (ONS) and represent the number of new SARS-CoV-2 positive test results by specimen date, each day from 1 March 2020 to 30 June 2020 (122 temporal observations). This is disaggregated by LTLA (380 of them), whose locations are provided in Figure \ref{fig:ltla.locations}. Note that there is great variation in the density of observation---akin to Scheme D in Section \ref{s:Simulations}---indicating that a spatial methodology may be especially effective. The trajectories for each of these locations is provided in Figure \ref{fig:covid.waves} (smoothed and normalised for comparison). Moreover, looking at the curves by region, it is clear there is some spatial correlation, with London being a much earlier and sharper peak than the East Midlands, for example.

\subsection{The Distance Metric}

\begin{figure}
\centering
\includegraphics[width=0.7\textwidth]{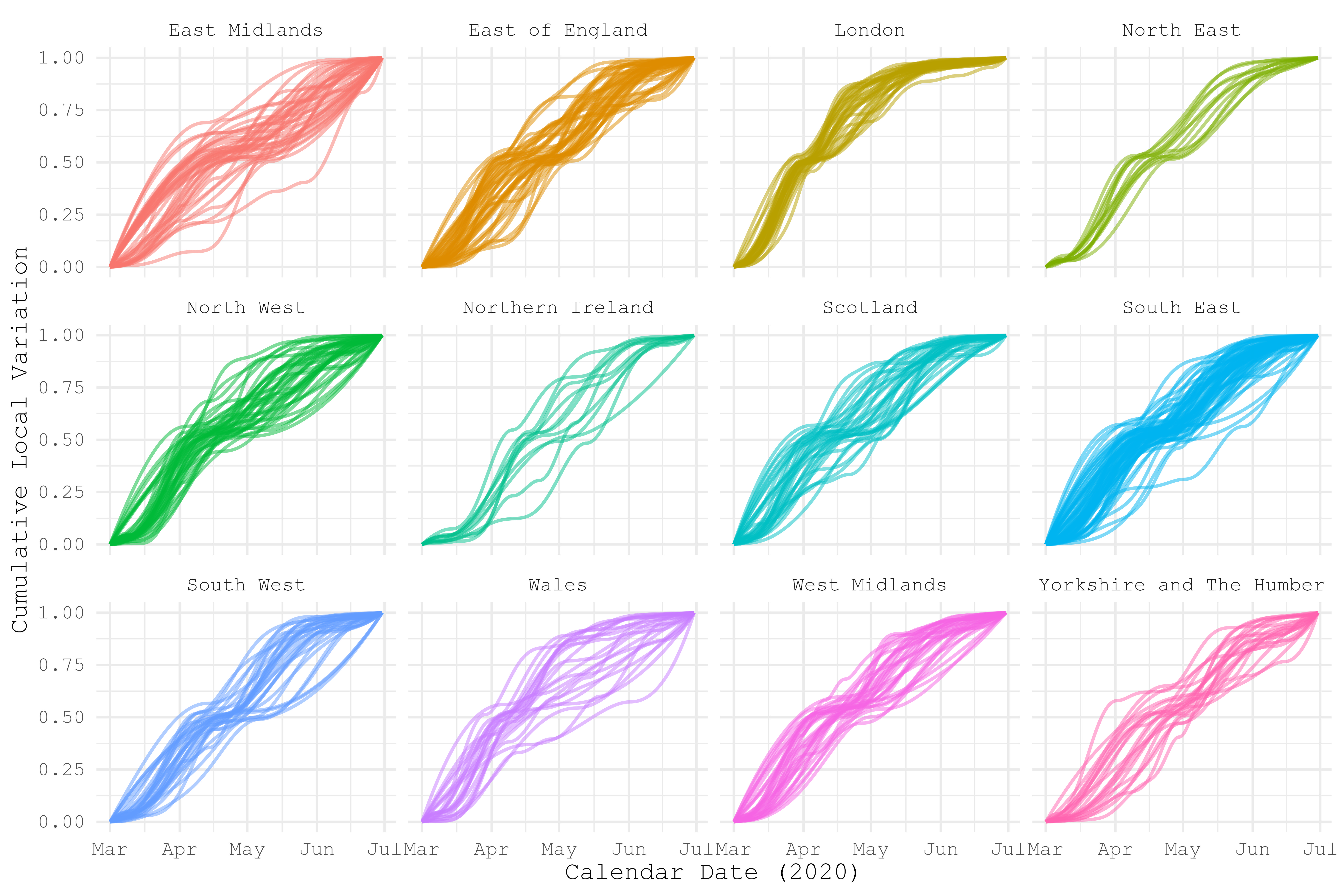}
\caption{Estimated cumulative local variation curves $\hat{\lambda}_i$. The facets are the nine regions of England, Wales, Scotland and Northern Ireland. For a geographic reference, see Figure \ref{fig:ltla.locations}.}
\label{fig:local.variation}
\end{figure}

To understand the phase variation, the data will be registered by LVA. First, this involves the estimation of the cumulative incidence variation, $\hat{\lambda}_i$, which are given in Figure \ref{fig:local.variation}. Note that these are calculated ignorant of any spatial correlation structure. Disaggregated by region, one can easily identify the spatial variation previously mentioned, with London peaking sharp and early compared to the East Midlands' wider and slower curve: the distribution of the incidence variation for London is more concentrated on a small interval when compared to the more uniform distribution of the East Midlands.

These estimates $\hat{\lambda}_i$ are then pairwise compared to produce an estimate of the variogram. The question remains though as to which distance metric to choose. When the Matérn variogram model is fitted by iteratively reweighted least squares with the geodetic distance, the estimated semisill, $\hat{\sigma}^2$, was less than $0.01\%$ of the estimated nugget, $\hat{\iota}$. This implies a lack of spatial correlation structure relative to inherent variation according a very general class of variograms, but restricted to a function of this geodetic distance. The idea that there is little covariance structure seems implausible, given the evident spatial nature of disease dynamic spread, so it may be concluded that geodetic distance is not a good parametrisation of this structure.

\begin{figure}
\centering
\includegraphics[width=0.7\textwidth]{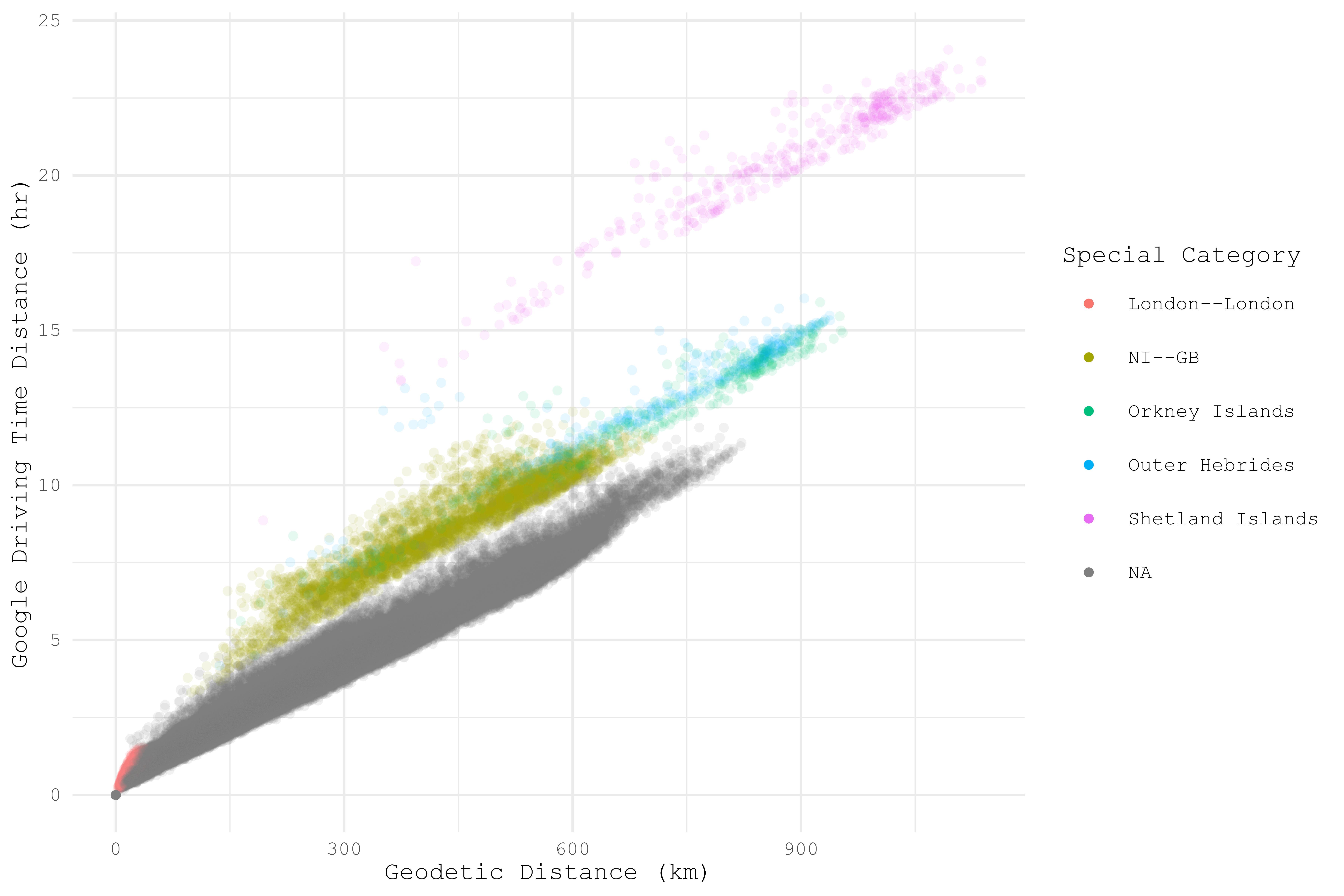}
\caption{Scatter plot of the Euclidean geodesic distances and driving time distances, coloured to highlight features of interest. In red are the London to London distances; note that the rate of growth of driving time increases faster in London than in general for pairs of locations, since the average driving speed in London is less than the rest of the country. For reference, the apparent lower bounding line of the scatter plot is around 60mph; the national speed limit on dual carriage ways is 70mph. In chartreuse are the pairs of locations where exactly one of them is in Northern Ireland, requiring the crossing of the Irish Sea. Similarly, in green, blue and magenta are pairs where one of which is the Orkneys, the Outer Hebrides and the Shetlands, respectively---all requiring crossings of bodies of water. These four locations account for the majority of divergence from the main area of the scatter plot.}
\label{fig:euclid.v.drivetime}
\end{figure}

A different distance which may elucidate the spatial correlation structure is one related to travel time between locations, a proxy for the rate at which populations in two locations may mix socially. The Google Routes API\footnote{See: \texttt{https://developers.google.com/maps/documentation/routes}.} was used to estimate average driving time between every (directed) pair of LTLA centroids. In general there is naturally a strong positive correlation between this and the geodesic distance on a sphere, but one can quickly see substantial divergences. Figure \ref{fig:euclid.v.drivetime} provides a scatter plot of the geodesic and drive time distances (appropriately modified; see below) for the pairs of observation locations. The first issue with the driving time, however, is that the raw values are neither symmetric nor obey the triangle inequality. To solve this, the average of the distance matrix and its transpose was taken to ensure symmetry, and for every $i,k,\ell\in[n]$ if $d_{k\ell}>d_{i_k}+d_{i\ell}$, then $d_{k\ell}$ was redefined as the right-hand sum, and this was iterated until no such breaches of the triangle equality occurred. The resulting distance on the set of 380 LTLAs is therefore a proper distance.

\begin{figure}
\centering
\includegraphics[width=0.7\textwidth]{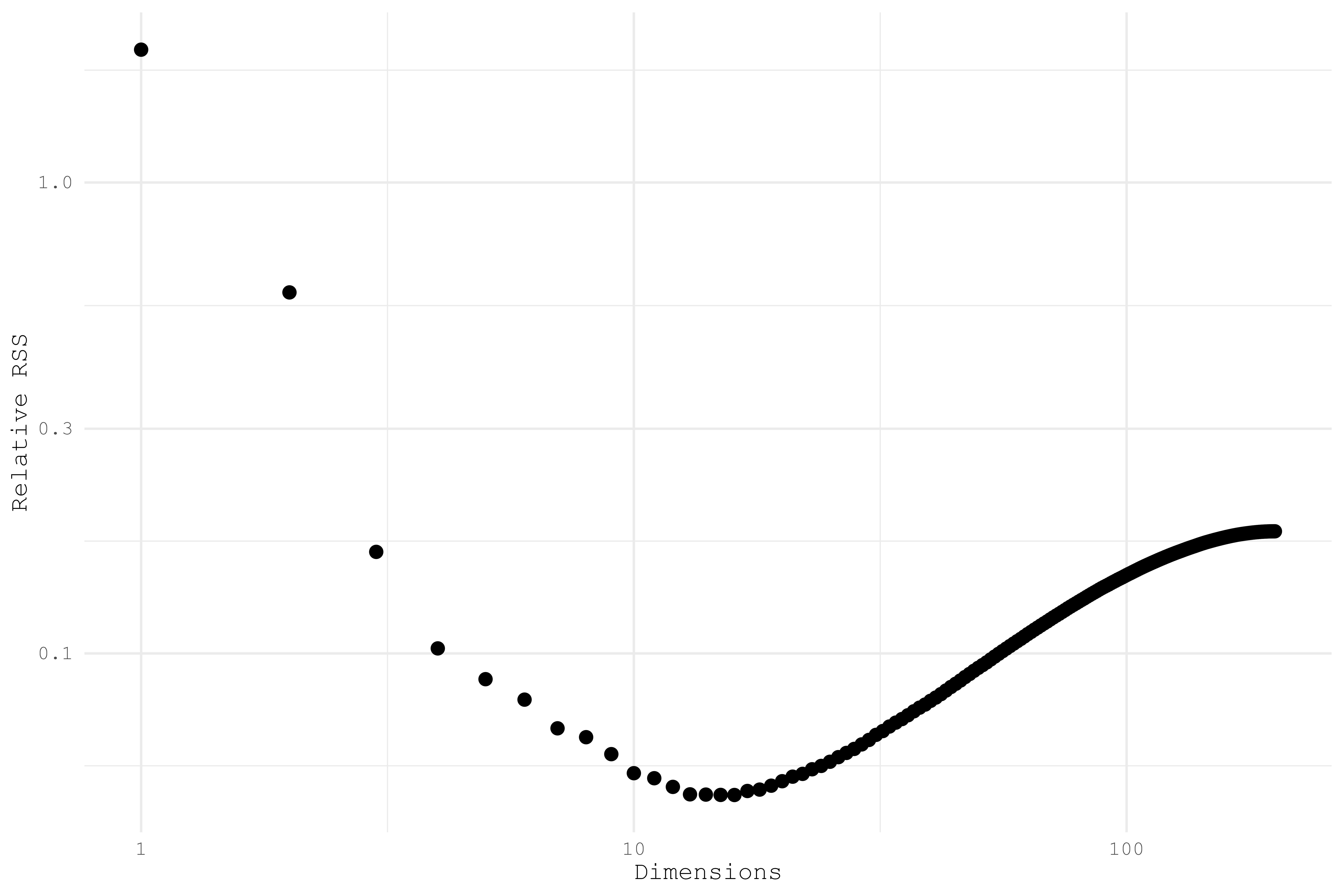}
\caption{Plot of the RSS in the regression (without intercept) of the drive times on the Euclidean embeddings by dimensionality, relative to the same on the geodesic distance. Note that in just two dimensions---the same dimensionality as the space in which the geodesic distance is approximately Euclidean---substantial progress is made in approximating the drive times.}
\label{fig:RSS.v.dim}
\end{figure}

\begin{figure}
\centering
\includegraphics[width=0.7\textwidth]{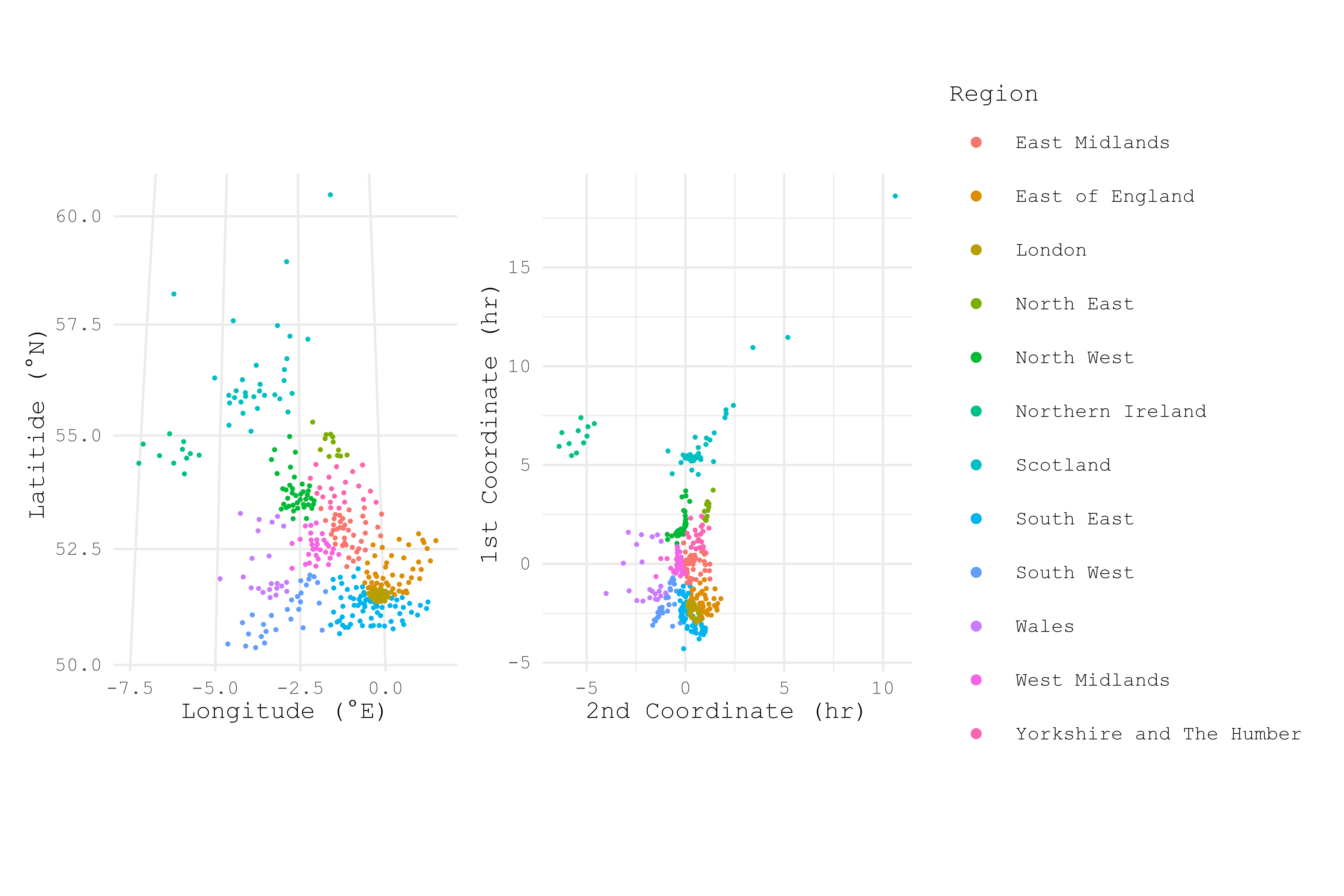}
\caption{Left: Centroids of the LTLAs on the Earth's surface. Right: The embedding of the locations in $\mathbb{R}^2$ such that Euclidean distance in the embedding approximates driving time between locations. The locations have been colour-coded by region for quick comparison. Note in particular the movement of Northern Ireland, the Shetlands, the Orkneys, the  Outer Hebrides and the Isle of Wight away from Great Britain the island. Also note that in the North of England, north--south connections are seen to be relatively faster than east--west connections; this largely corresponds to motorways.}
\label{fig:R2.rep.drivetime}
\end{figure}

There remains the issue, however, that this distance is not a Euclidean distance, and so it is not clear that the variogram family will be positive definite with respect to this distance (indeed, the Matérn family is not). The distance metric methodology considered above (Algorithm \ref{alg:dist}) was therefore applied. For each of the possible dimensions $p$ (in this application, $B$ had 200 positive eigenvalues out of 380), one can compute an RSS of the linear regression of the non-Euclidean distances $d_{ik}$ against the Euclidean distances $d^*_{ik}$ (with no intercept term), and compare it relative to the RSS of the same for the $d_{ik}$ against the (nearly-Euclidean) geodetic distances. This is the distance between matrices as in criteria (1) for choosing $p$. The output is given in Figure \ref{fig:RSS.v.dim}, which demonstrates the optimal dimensionality of the embedding space with this methodology is 16. For visualisation purposes, anything higher than three dimensions is of course not very feasible, but as seen in Figure \ref{fig:RSS.v.dim} even two dimensions is enough to pick up some characteristics of the difference between geodetic distance and driving times. The plot of LTLA locations in the approximately two-dimensional Euclidean space of the Earth's surface next to the plot of LTLA locations in the two-dimensional embedding that approximates travel time is given in Figure \ref{fig:R2.rep.drivetime}.

\begin{figure}
\centering
\includegraphics[width=0.7\textwidth]{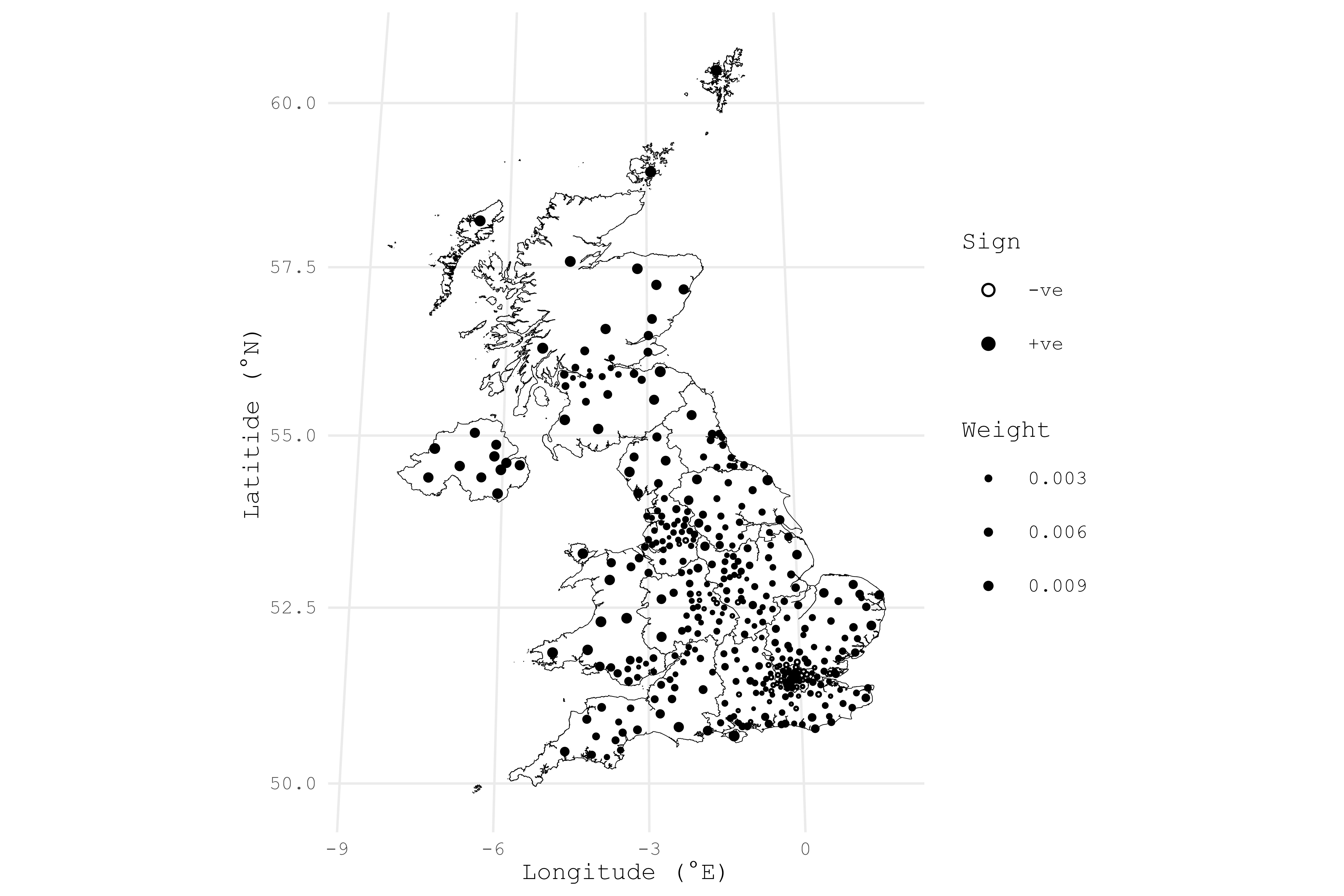}
\caption{Calculated weights in the averaging of the $\hat{\lambda}_i$ mapped by geographical location. Note the greater weighting in sparsely observed areas. (Note that sparse observations here is relative to the modified driving time distance, rather than geographical distance.)}
\label{fig:weights}
\end{figure}

Now that a Euclidean distance that approximates the driving time distance has been established, a new variogram can be estimated by iteratively reweighted least squares. For the Matérn family as a function of this 16-dimensional Euclidean distance (note that the Matérn family is valid in all dimensions), the parameter estimates are: $\hat{\iota}=0.00772$, $\hat{\sigma}^2=0.00860$, $\hat{\nu}=112$ and $\hat{\rho}=2330$. (Here the distance is measured in units approximately drive time seconds.) Note in particular that the semisill is indeed greater than the nugget, and so there is a substantial spatial effect. This spatial effect has a range (the distance at which 95\% of the maximal variation is achieved, the typical distance at which observations may be considered uncorrelated) of around 84 minutes (i.e. two locations an 84-minute drive away from each other have little to do with each other when it comes to the Covid-19 spatiotemporal dynamics). From this estimate of the variogram, it is straightforward to compute the weights used in the averaging of the $\hat{\lambda}_i$; these are mapped in Figure \ref{fig:weights}. Much like with the toy example in the introduction, areas sparsely observed are weighted higher than areas densely observed; although, it must be kept in mind that density of observation is now relative the the modified driving time distance, not geographical distance as the mapping would suggest. `Sparsely' observed areas may therefore now instead be thought of as `isolated' areas.

\subsection{The Warping Functions}

\begin{figure}
\centering
\includegraphics[width=0.7\textwidth]{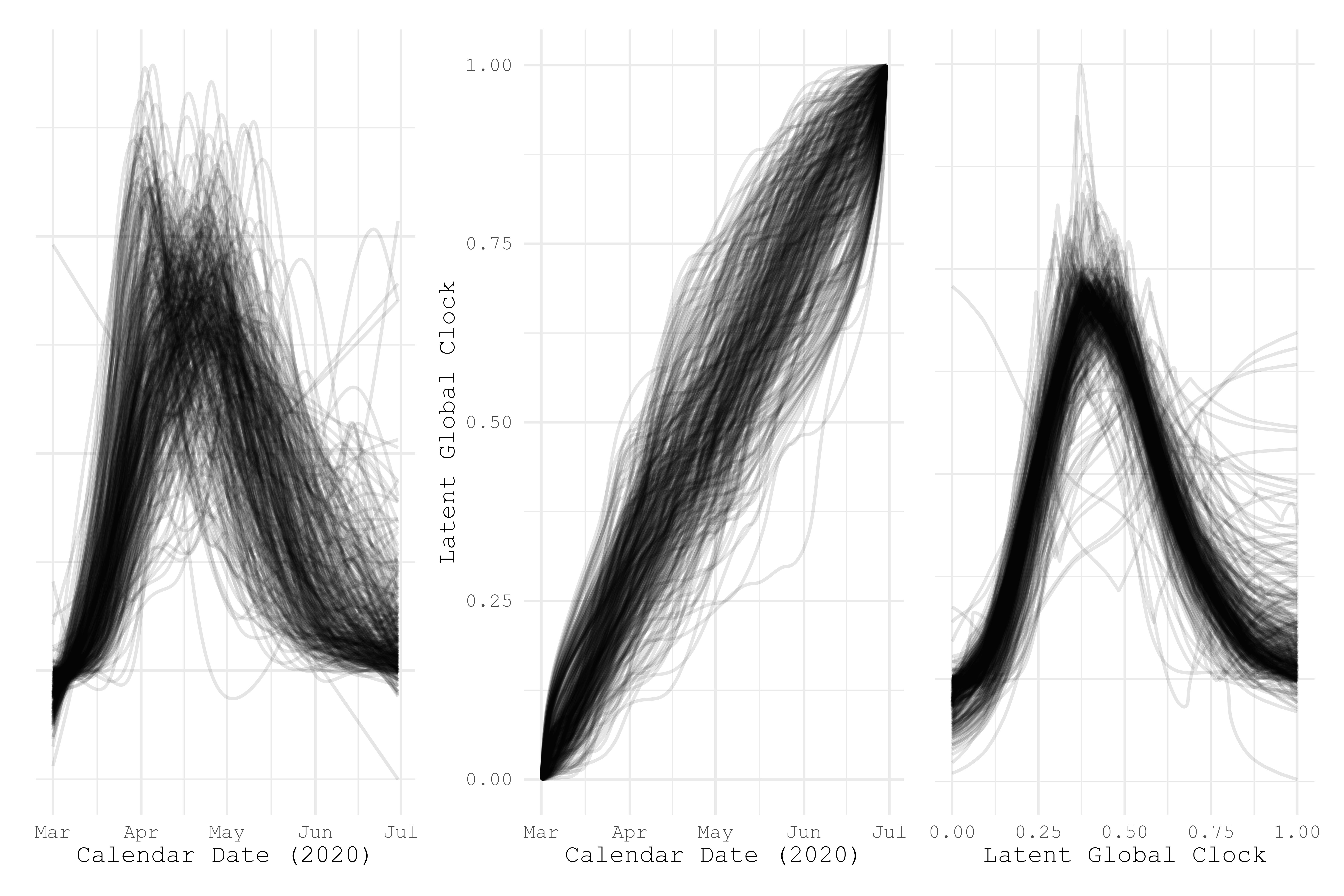}
\caption{Left: The unaligned but smoothed and normalised SARS-CoV-2 incidence trajectories $\hat{y}_i$ by LTLA in the UK. Centre: The estimated warping functions $\hat{h}_i^{-1}$, mapping recorded calendar time to a latent global clock under which features are aligned. Right: The estimated aligned curves (normalised) $\hat{x}_i$.}
\label{fig:unaligned.2.aligned}
\end{figure}

With these weightings, $\hat{\lambda}_\mu$ and thereby the $\hat{h}_i^{-1}$ are then estimated, along with the aligned $\hat{x}_i$. These are provided in Figure \ref{fig:unaligned.2.aligned}. These aligned curves can then be used for more parsimonious analyses of the development of Covid-19 in the UK, but it is the alignment processes themselves which are of interest here. These are analysed now.

\begin{figure}
\centering
\includegraphics[width=0.7\textwidth]{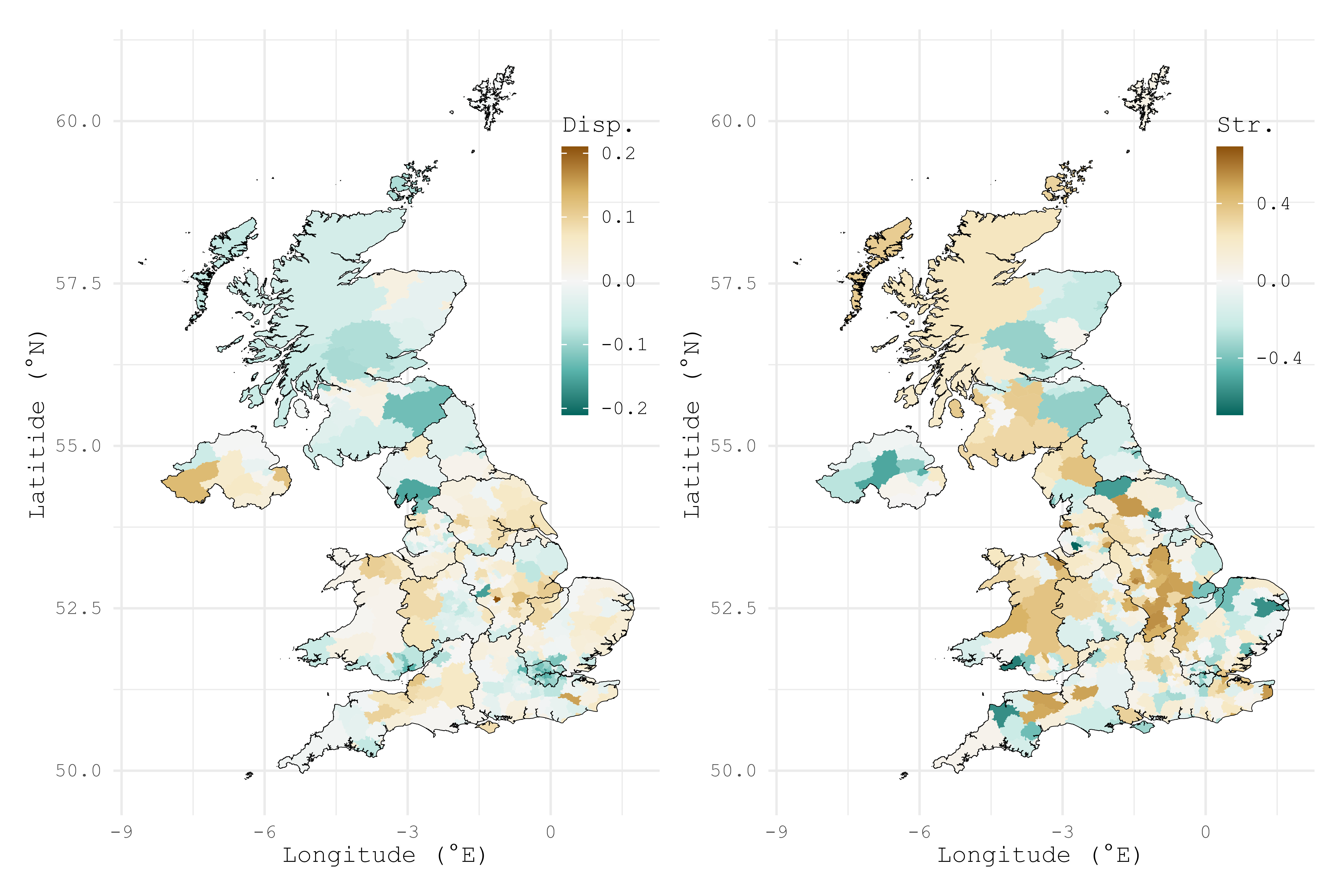}
\caption{Left: Choropleth of the displacement functional. Note the relative earliness of the wave in London vs the relative lateness in the East Midlands. Right: Choropleth of the stretch functional. Note the relative sharpness of the wave in London vs the relative flatness in the East Midlands.}
\label{fig:choropleth.functionals}
\end{figure}

To analyse the estimates $\hat{h}_i^{-1}$, two functionals are defined. First, there is displacement:
\begin{equation}
\Delta_i:=\int_0^1 t\mathrm{d}H_i^{-1}(t) - \frac{1}{2}.
\end{equation}
Considering $H_i^{-1}$ as a distribution, this is the expectation of the latent clock with respect to the measured clock, minus the expectation of $\mathrm{id}$. Negative values therefore represent a displacement of the curves earlier, and positive values represent a displacement of the curves later. A choropleth map (where each areal unit is coloured to represent some level in the area) of the estimated displacements by LTLA is provided in Figure \ref{fig:choropleth.functionals}.

Second, there is stretch:
\begin{equation}
\Upsilon_i:=\log\left(12\int_0^1 \left(t-\Delta_i-\frac{1}{2}\right)^2\mathrm{d}H_i^{-1}(t)\right).
\end{equation}
This is the variance of the latent clock with respect to the measured clock, divided by the variance of $\mathrm{id}$ and the logarithm taken (so that zero is the central tendency). Negative values therefore represent relatively sharp peaks, and positive values represent relatively wide peaks. A choropleth map of the estimated stretches by LTLA is also provided in Figure \ref{fig:choropleth.functionals}.

\begin{figure}
\includegraphics[width=0.7\textwidth]{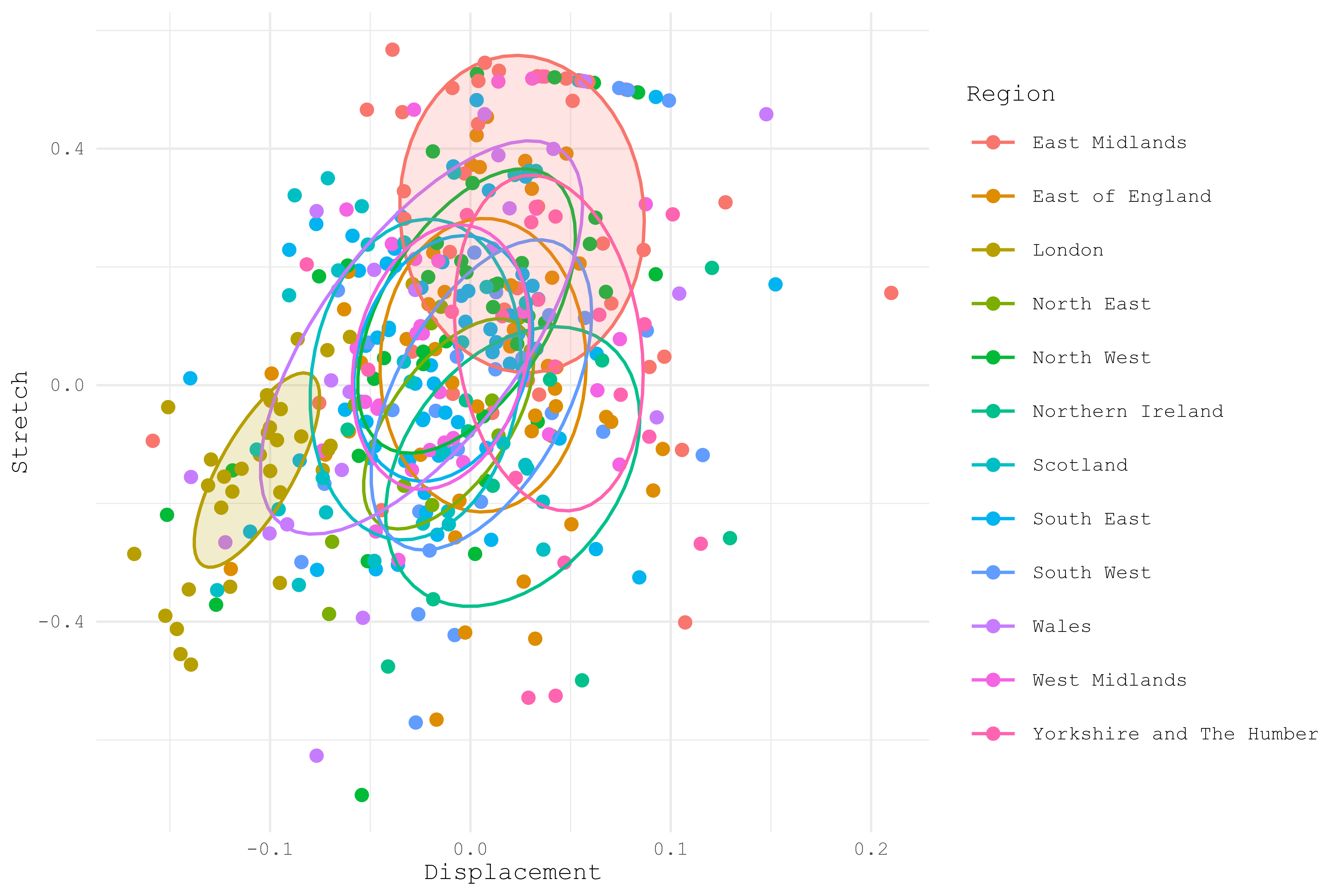}
\caption{Scatter plot of the displacement and stretch functionals for each LTLA, with 50\%-level ellipses for each region. London and the East Midlands have their 50\%-level ellipses highlighted.}
\label{fig:str.v.disp}
\end{figure}

These choropleths now quantiatively demonstrate the qualitative analysis of the curves: London had early and sharp peaks, while the East Midlands peaked later and wider. If left in any doubt, Figure \ref{fig:str.v.disp} provides a scatter plot of these functionals with 50\%-level ellipses for each region, clearly showing a distinction between London and the East Midlands. However, it is also of interest to see that there are a range of times and shapes across the UK, indicating that disease dynamics really did differ depending on where in the country is being considered.

Furthermore, these scalar functionals of the global warping functions can now be input into traditional geospatial statistics methodologies, such as kriging (prediction at intermediary locations), spatial clustering (partitioning of the map into regions of differing behaviour) and wombling (finding contours of sudden change). For more information, see \cite{cressie2015statistics} and/or \cite{gelfand2010handbook}.

\section{Discussion}

To summarise, to understand the phase variation and hence timing of the first Covid-19 waves in the United Kingdom across geography, two methodologies have been developed; the first to approximate a non-Euclidean distance by a Euclidean one, which then mathematically enables the second to estimate the local phase variations taking into account the spatial correlation structure. This has allowed a more granular understanding of the wave timings across the United Kingdom.

For the Covid-19 data, it has been shown that the use of the driving time distance is more appropriate than a geodetic distance, since it recovers a much greater spatial correlation structure relative to the inherent variation. The Covid-19 data has demonstrated that the parametrisation of a spatial model in terms of a distance metric may be quite sensitive to the choice of distance, and that the interpretation of this distance can be in itself statistically interesting (see, for example, the range of 1.5 hour driving hours for the spatial correlation model). The freedom the distance methodology of Section \ref{s:DistanceMetric} provides in the choice of distance metric opens up other opportunities. The methodology, inspired by multidimensional scaling, has applications in geospatial statistical analysis much more broadly: it is not uncommon to be extrinsically interested in a non-Euclidean distance; meanwhile, the theoretical frameworks established often rely on the use of a Euclidean distance. While some authors overlook this issue, and others work to mitigate it on a case-by-case basis, this methodology provides a systematic approach which may perform well enough in a variety of applications. There are still theoretical improvements that could be made since there is no guarantee of the approximation in a $p$-dimensional Euclidean space is optimal, although the computational simplicity of the algorithm is advantageous. For driving times vs geodetic distance in evaluating Covid-19 cases, over 75\% of the difference between the two could be overcome by a Euclidean approximation from this methodology.

Having established the distance metric and associated spatial correlation model, the Covid-19 waves could be registered in a spatially aware way, which has been demonstrated to reduce the mean squared error of resulting estimates of local phase variation when compared to the existing non-spatial methodology. The spatially aware functional registration methodology is itself applicable independently of the Euclidean distance approximation methodology, and the underlying ideas involved are applicable for various registration methodologies (see Appendix A), which therefore allows for the use of different assumptions on the phase variation model; moreover, it is plausible to extend this methodology to include, for example, covariates of interest.

The resulting estimates of the local phase variation have then been analysed by taking functionals to understand the implied phase variation across geography via a low-dimensional representation of the essential characteristics of the phase variation. This allows the visualisation of the phase variation (living in an infinite-dimensional functional space) across geography, by picking out its essential components. The warping estimates themselves may be useful in further analysis, by converting calendar time-dependent covariates to the latent global clock, so that, for example, a national lockdown measure is understood locally, relative to its timing in the Covid-19 wave.

\bibliographystyle{Chicago}

\bibliography{ref}

\appendix

\input{sup_appendix_a_arxiv.tex}

\input{sup_appendix_b_arxiv.tex}
\end{document}

%% file: sup_appendix_a_arxiv.tex
\section{Other Registration Methodologies}

The choice to adapt the local variation analysis (LVA) registration methodology to a spatial model was convenient both computationally and theoretically, but it was not unique. For comparison, the same heuristic has been applied to the pairwise curve synchronisation (PCS) approach of Tang and M\"uller \citep{tang2008pairwise} and the Fisher--Rao geometry (FRG) approach of Srivastava et al. \citep{srivastava2011registration}, although both these have their challenges.

In particular, for the spatial PCS approach, a variogram must be estimated for the pairwise warping functions at each central location, since the spatial functional field is not stationary due to the correlation in the warping process. This results in $O(n)$ optimisations to be made instead of the $O(1)$ optimisation steps in the spatial LVA methodology. Moreover, as demonstrated by Chakraborty and Panaretos \citep{chakraborty2021functional}, this methodology does not perform under the given assumptions as well as the LVA and FRG methodologies (in the non-spatial case). On the other hand, for the spatial FRG approach, the number of optimisations is no longer due to the estimation of multiple variograms, but in computing the Fisher--Rao distance metrics themselves, which are in fact $O(n^2)$. This complexity is reflected in the average computation time for each of the methodologies in the simulation setup: 1.53 seconds for LVA, 115 seconds for PCS, and 590 seconds for FRG. It may be the case that computational efficiencies can be found for both computation of the fitted variograms and the Fisher--Rao distance metrics, but for the purposes of simulation and thereby validating the spatialisation of the methodology, it was only reasonable to do so for the spatial vs non-spatial LVA methodologies; however, small-sample results suggest that spatial LVA does perform better than spatial PCS and spatial FRG, the former mirroring the results of Chakraborty and Panaretos \citep{chakraborty2021functional} (as well as some theoretical diffifculties) and the latter reflecting theoretical difficulties in spatialising the FRG approach. (In particular, for the PCS approach there is a violation of the stationarity assumption, and the FRG approach, due to the non-linearity in the Fisher--Rao geometry, the estimated weights will only serve as approximations to the optimal weights, which are somewhat intractable.)

\subsection{Pariwise Curve Syncrhonisation (PCS)}

A spatial variation of the PCS methodology of Tang and M\"uller \citep{tang2008pairwise} is provided below:
\begin{enumerate}
\item Estimate the $Z_i:=\mu\circ H_i^{-1}$ by normalising the estimated curves $\hat{Y}_i$.
\item Estimate the pairwise warping functions $G_{ik}:=H_i\circ H_k^{-1}$ by a minimisation problem, e.g.:
\begin{equation}
\hat{G}_{ik} := \underset{\tilde{g}\in\hat{\mathcal{W}}}{\argmin}\left[||\hat{Z}_i\circ\tilde{g} - \hat{Z}_k||^2 + \rho ||\tilde{g}-\mathrm{id}||^2\right],
\end{equation}
where $\rho$ is some regularisation parameter and $\hat{\mathcal{W}}$ is a finite-dimensional approximation of $\mathcal{W}$.
\item Estimate the variograms $\gamma^{(i)}$ based on the data $\left(d_{k\ell}, ||\hat{G}_{ki}-\hat{G}_{\ell i}||^2\right)_{k,\ell\neq i}$.
\item Estimate the approximately optimal weights:
\begin{equation}
\hat{w}^{(i)}:=\frac{\left(\hat{C}^{(i)}\right)^{-1}1_{n-1}}{1_{n-1}^\top \left(\hat{C}^{(i)}\right)^{-1} 1_{n-1}},\quad \hat{C}^{(i)}_{k\ell}:= 2\hat{\gamma}^{(i)}(\infty) - 2\hat{\gamma}^{(i)}(d_{k\ell}).
\end{equation}
\item Estimate the global warping functions:
\begin{equation}
\hat{H}_i^{-1} := \frac{1}{n-1}\sum\limits_{k\neq i} \hat{w}^{(i)}_k \hat{G}_{ki}.
\end{equation}
\end{enumerate}

\subsection{Fisher--Rao Geometry (FRG)}

A spatial variation of the FRG methodology of Srivastava et al. \citep{srivastava2011registration} is provided below:
\begin{enumerate}
\item Estimate the full curves $Y_i$ and thereby their full orbits $\mathcal{O}_i=Y_i\circ\mathcal{W}$.
\item Estimate the variogram $\gamma^{(1)}$ from the data $\left(d_{ik}, d^2_{\mathcal{F}/\mathcal{W}}(\hat{\mathcal{O}}_i,\hat{\mathcal{O}}_k)\right)_{i\neq k}$, where $d_{\mathcal{F}/\mathcal{W}}$ is the Fisher--Rao distance on the orbits of absolutely continuous functions $\mathcal{F}/\mathcal{W}$.
\item Estimate the approximately optimal weights:
\begin{equation}
\hat{w}^{(1)} := \frac{\left(\hat{C}^{(1)}\right)^{-1} 1_n}{1_n^\top \left(\hat{C}^{(1)}\right)^{-1}  1_n},\quad \hat{C}^{(1)}_{ik} :=2\hat{\gamma}^{(1)}(\infty) - 2\hat{\gamma}^{(1)}(d_{ik}).
\end{equation}
\item Estimate the template orbit by a weighted Karcher mean:
\begin{equation}
\hat{\mathcal{K}}:=\underset{\tilde{\mathcal{O}}\in \mathcal{F}/\mathcal{W}}{\argmin}\sum\limits_{i=1}^n \hat{w}^{(1)}_i d_{\mathcal{F}/\mathcal{W}}(\tilde{\mathcal{O}}, \mathcal{O}_i).
\end{equation}
\item Pick an element $\hat{K}^\dagger\in\hat{\mathcal{K}}$, and compute the following warpings:
\begin{equation}
\hat{\theta}^\dagger_i:=\underset{\tilde{\theta}\in\mathcal{W}}{\argmin} d_\mathcal{F}\left(\hat{K}^\dagger,\hat{Y}_i\circ\tilde{\theta}\right),
\end{equation}
where $d_\mathcal{F}$ is the Fisher--Rao distance on absolutely continuous functions $\mathcal{F}$.
\item Estimate the variogram $\gamma^{(2)}$ from the data $\left(d_{ik},d^2_\mathcal{W}(\hat{\theta}_i^\dagger,\hat{\theta}_k^\dagger)\right)_{i\neq k}$, where $d_\mathcal{W}$ is the Fisher--Rao distance on warping functions $\mathcal{W}$.
\item Estimate the approximately optimal weights:
\begin{equation}
\hat{w}^{(2)} :=\frac{\left(\hat{C}^{(2}\right)^{-1} 1_n}{1_n^\top \left(\hat{C}^{(2)}\right)^{-1}  1_n},\quad \hat{C}^{(2)}_{ik}:= 2\hat{\gamma}^{(2)}(\infty) - 2\hat{\gamma}^{(2)}(d_{ik}).
\end{equation}
\item Estimate the weighted Karcher mean of the warpings:
\begin{equation}
\hat{\kappa}:=\underset{\tilde{\theta}\in\mathcal{W}}{\argmin}\sum\limits_{i=1}^n \hat{w}^{(2)}_i d_\mathcal{W}(\tilde{\theta},\hat{\theta}_i^\dagger),
\end{equation}
and thereby compute the weighted centre of $\hat{K}$ with respect to the $\hat{Y}_i$ and $\hat{w}^{(2)}$ as $\hat{K}:=\hat{K}^\dagger\circ\hat{\kappa}^{-1}$.
\item Estimate the global warping functions as:
\begin{equation}
\hat{H}_i := \underset{\tilde{\theta}\in\mathcal{W}}{\argmin} d_\mathcal{F} (\hat{K}, \hat{Y}_i\circ\tilde{\theta}).
\end{equation}
\end{enumerate}

%% file: sup_appendix_b_arxiv.tex
\section{Simulation Particulars}

\subsection{The Observation Schemata}
The observation locations were produced (pseudo-)randomly as:
\begin{description}
    \item[Scheme A] A $6\times6$ square grid, with horizontal and vertical coordinates at $\frac{2\alpha - 1}{12}$ for $\alpha\in[6]$.
    \item[Scheme B] Uniformly and independently distributed on $[0,1]^2$.
    \item[Scheme C] Independently as a function of $R\sim |\mathsf{Normal}(0,0.04)|$ and $\Theta\sim \mathsf{Uniform}[0,2\pi]$, with horizontal component $0.5+R\cos\Theta$ and vertical component $0.5+R\sin\Theta$ (i.e. concentrated at a central point).
    \item[Scheme D] Independently of three forms:
    \begin{itemize}
	\item one point at $(0.2,0.2)$;
	\item ten points distributed as bivariate normal: \[\mathsf{Normal}_2\left(\begin{pmatrix}0.8\\0.4\end{pmatrix},\begin{pmatrix}0.005&0.005\\-0.005&0.04\end{pmatrix}\right);\]
	\item twenty-five points distributed as bivariate normal: \[\mathsf{Normal}_2\left(\begin{pmatrix}0.3\\0.7\end{pmatrix},\begin{pmatrix}0.008&-0.005\\-0.005&0.008\end{pmatrix}\right).\]
    \end{itemize}
\end{description}

\begin{figure}
        \centering
        \includegraphics[width=0.7\textwidth]{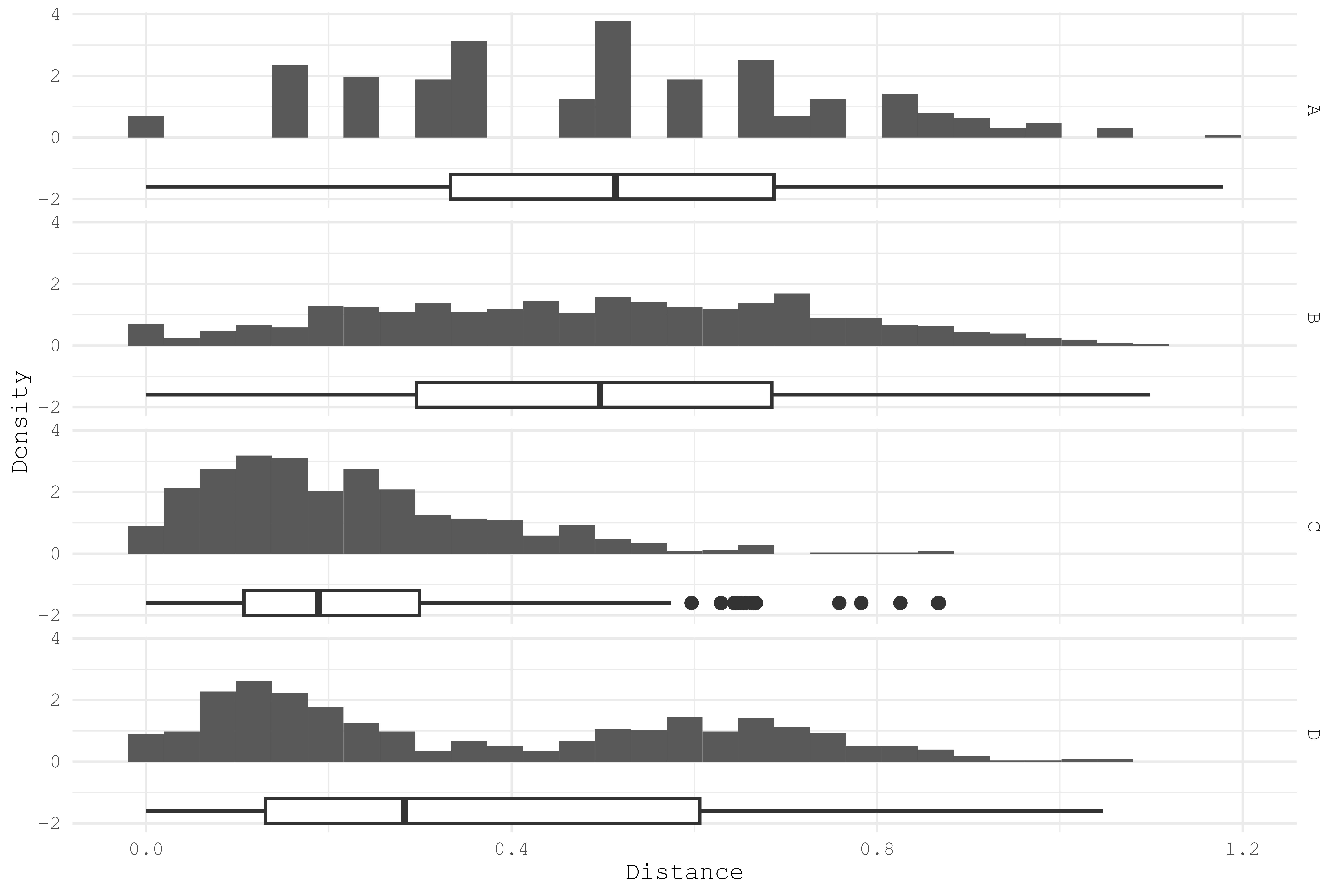}
        \caption{Histograms and boxplots for the pairwise distances of the points produced via the four spatial observation schemata as described in text.}
        \label{fig:sim.dists}
\end{figure}

The pairwise distances between observations under each of the schemata were calculated, with the results provided in Figure \ref{fig:sim.dists}. These demonstrate that the distribution of the pairwise distances are broadly comparable, allowing for the same isotropic covariance structure to be used comparably across the four schemata.

\subsection{The Generation of the Warpings}

This covariance structure fed into the generation of the $h_i^{-1}$. In particular, two (pseudo-)random variables were simulated:
\begin{equation}
    \zeta_j \overset{\mathrm{iid}}{\sim} \mathsf{Normal}_n\left( 0_n, C \right),\quad j \in [2],
\end{equation}
where $0_n\in\mathbb{R}^n$ represents a length-$n$ vector of zeros, and $C$ is the $n\times n$ matrix of covariances $(C(d_{ik}))_{i,k\in[n]}$. Writing $\zeta_{ji}^* := 0.25\Phi\left(\zeta_{ji}\right)-0.125\in[-0.125,0.125]$ where $\Phi$ is the cumulative distribution function of a standard Gaussian, these then produced the $h_i^{-1}$ via:
\begin{equation}
    h_i^{-1}(t) := \begin{cases}
        \frac{0.25+\zeta_{1i}^*}{0.25-\zeta_{1i}^*}t,& 0 \leq t < 0.25 - \zeta_{1i}^*\\
        \frac{0.5 + \zeta_{2i}^* - \zeta_{1i}^*}{0.5 - \zeta_{2i}^* + \zeta_{1i}^*}(t - 0.25 + \zeta_{1i}^*) + 0.25 + \zeta_{1i}^*,& 0.25 - \zeta_{1i}^* \leq t < 0.75 - \zeta_{2i}^*\\
        \frac{0.25 - \zeta_{2i}^*}{0.25 + \zeta_{2i}^*} (t - 0.75 - \zeta_{2i}^*) + 0.75 - \zeta_{2i}^*,& 0.75 - \zeta_{2i}^* \leq t \leq 1
    \end{cases},
\end{equation}
that is the linear interpolation of the ordinates $(0, 0.25 - \zeta_{1i}^*, 0.75 - \zeta_{2i}^*, 1)$ and the abscissae $(0, 0.25 + \zeta_{1i}^*, 0.75 + \zeta_{2i}^*, 1)$.

\subsection{The Simulation Results}

The estimates of the MSEs with 95\% confidence intervals after 5000 simulations are provided in detail in Table \ref{tab:sim.results}.
\begin{table}
    \centering
    \begin{footnotesize}
    \begin{tabular}{cc|cc|cc}
         & & \multicolumn{2}{|c|}{Non-Spatial} & \multicolumn{2}{|c}{Spatial} \\
         Scheme & $\psi$ & Estimate & $\pm$ & Estimate & $\pm$ \\\hline
\multirow{4}{*}{A} & 0.03 & 0.5366 & 0.0093 & 0.5399 & 0.0093 \\
 & 0.1 & 0.8197 & 0.0176 & 0.8074 & 0.0171 \\
 & 0.3 & 2.4934 & 0.0626 & 2.1611 & 0.0548 \\
 & 1 & 5.8964 & 0.1371 & 5.0152 & 0.1217 \\ \hline
\multirow{4}{*}{B} & 0.03 & 0.5967 & 0.0107 & 0.5944 & 0.0104 \\
 & 0.1 & 1.0768 & 0.0236 & 0.9647 & 0.0209 \\
 & 0.3 & 3.0768 & 0.0746 & 2.7018 & 0.0661 \\
 & 1 & 6.7300 & 0.1498 & 5.9732 & 0.1369 \\ \hline
\multirow{4}{*}{C} & 0.03 & 0.9887 & 0.0228 & 0.7999 & 0.0181 \\
 & 0.1 & 2.8329 & 0.0724 & 1.7838 & 0.0491 \\
 & 0.3 & 5.8810 & 0.1416 & 3.7214 & 0.1027 \\
 & 1 & 8.8845 & 0.1996 & 6.7073 & 0.1624 \\ \hline
\multirow{4}{*}{D} & 0.03 & 0.9158 & 0.0268 & 0.8021 & 0.0231 \\
 & 0.1 & 2.0745 & 0.0691 & 1.3408 & 0.0462 \\
 & 0.3 & 5.1652 & 0.1603 & 3.4631 & 0.1253 \\
 & 1 & 9.1104 & 0.2545 & 7.7934 & 0.2316 \\ \hline
    \end{tabular}
    \end{footnotesize}
    \caption{Simulation results, all $\times 10^{-3}$. For each of the four schemata, for each of the covariance structures, and for each spatial and non-spatial versions of the methodology, the simulated mean squared error, $\frac{1}{n}\sum_{i=1}^n\mathbb{E}\left|\left|\hat{H}^{-1}_i - H^{-1}_i\right|\right|^2$, along with the radius of the 95\% confidence interval.}
    \label{tab:sim.results}
\end{table}